\documentclass{jfm}

\usepackage{graphicx}
\usepackage{color}
\usepackage{natbib}
\graphicspath{{Figs/}}
\ifCUPmtlplainloaded \else
  \checkfont{eurm10}
  \iffontfound
    \IfFileExists{upmath.sty}
      {\typeout{^^JFound AMS Euler Roman fonts on the system,
                   using the 'upmath' package.^^J}%
       \usepackage{upmath}}
      {\typeout{^^JFound AMS Euler Roman fonts on the system, but you
                   dont seem to have the}%
       \typeout{'upmath' package installed. JFM.cls can take advantage
                 of these fonts,^^Jif you use 'upmath' package.^^J}%
       \providecommand\upi{\pi}%
      }
  \else
    \providecommand\upi{\pi}%
  \fi
\fi

\ifCUPmtlplainloaded \else
  \checkfont{msam10}
  \iffontfound
    \IfFileExists{amssymb.sty}
      {\typeout{^^JFound AMS Symbol fonts on the system, using the
                'amssymb' package.^^J}%
       \usepackage{amssymb}%

      }{}
  \fi
\fi

\ifCUPmtlplainloaded \else
  \IfFileExists{amsbsy.sty}
    {\typeout{^^JFound the 'amsbsy' package on the system, using it.^^J}%
     \usepackage{amsbsy}}
    {}
\fi

\newsavebox{\astrutbox}
\sbox{\astrutbox}{\rule[-5pt]{0pt}{20pt}}

\def\upi{\pi}
\def\kv{{\mathbf k}}  
\def\xv{{\mathbf x}}  
\def\xiv{\mbox{\boldmath $\xi$}}
\def\chiv{{\mbox{\boldmath $\chi$}}} 
\def\zetav{{\mbox{\boldmath $\zeta$}}} 

\title[Universality of Sea Wave Growth]{Universality of Sea Wave Growth and Its Physical Roots}

\author[V.~E.~Zakharov, S.~I.~Badulin, P.~A.~Hwang and G.~Caulliez]%
{Vladimir E. Zakharov$^{1-5}$
,\ns
Sergei~I.~Badulin$^{2,3}$,\thanks{Email address for correspondence: badulin@ioran.ru}\break
 Paul~A.~Hwang$^6$ and Guillemette Caulliez$^7$}

\affiliation{$^1$University of Arizona, Tuscon, USA\\
[\affilskip] $2$ Laboratory of Nonlinear Wave Processes, Novosibirsk State University, Russia\\
[\affilskip] $^3$P.P.Shirshov Institute of Oceanology of Russian
Academy of Sciences Moscow, Russia \\
[\affilskip] $^4$P.N. Lebedev Physical Institute of Russian Academy of Sciences\\
[\affilskip] $^5$Waves and Solitons LLC, Phoenix, Arizona, USA \\
[\affilskip] $^6$Remote Sensing Division, Naval Research Laboratory, Washington, D.C.\\
[\affilskip] $^7$Aix-Marseille Universit\'e, Universit\'e de Toulon, CNRS/INSU, IRD, MIO, UM 110, 13288, Marseille, Cedex 09, France}

\pubyear{2014}
\volume{000}
\pagerange{000--000}
\date{22 November 2014; revised ?; accepted ?. - To be entered by editorial office}
\begin{document}

\maketitle

\begin{abstract}
Modern day studies of wind-driven sea waves are usually focused on wind forcing rather than on the effect of resonant nonlinear wave interactions. The authors assume that these effects are dominating and propose a simple relationship between instant wave steepness and time or fetch of wave development expressed in wave periods or lengths. This law does not contain wind speed explicitly and relies upon this  asymptotic theory. The validity of this law is illustrated by results of numerical simulations, in situ measurements of growing wind seas and wind wave tank experiments. The impact of the new vision of sea wave physics is discussed in the context of conventional approaches to wave modeling and forecasting.
\end{abstract}

\begin{keywords}
Authors should not enter keywords on the manuscript, as these must be chosen by the author during the online submission process and will then be added during the typesetting process (see http://journals.cambridge.org/data/\linebreak[3]relatedlink/jfm-\linebreak[3]keywords.pdf for the full list)
\end{keywords}

\section{Introduction}

Wind-driven waves are usually seen as well-understood phenomenon of which the physics looks `self-evident': waves are growing due to wind and dissipate due to  wave breaking. In this scenario effects of resonant nonlinear wave interactions are  considered  just as minor corrections making some redistribution of wave energy along the  spectrum. This `common-sense' understanding of sea wave physics finds its reflection in conventional scaling of wave growth by wind speed \citep[e.g.][]{SverdrupMunk1947,Kitai62} and in attempts to find features of wave growth universality in terms of such scaling. Non-dimensional wave height variance $\varepsilon$ (wave energy in the wave community lingo) defined by
\begin{equation}\label{eq:EnND}
\varepsilon=\frac{Eg^2}{U_h^4}
\end{equation}
and the non-dimensional characteristic wave frequency
\begin{equation}\label{eq:FreqND}
\sigma=\frac{\tilde \omega U_h}{g}
\end{equation}
with $g$ - gravity acceleration, $U_h$ the wind speed at a reference height $h$ or  the friction velocity $u^*$ (for a constant flux boundary layer $u^*=(<U'W'>)^{1/2}$), and $E=\langle |\eta|^2\rangle $ -- wave height variance, are widely used in experimental and numerical studies. The characteristic wave frequency $\tilde\omega$ in (\ref{eq:FreqND}) can be defined as the mean-over-spectrum $\omega_m$, the zero-crossing $\omega_z$ or the spectral peak frequency $\omega_p$. Below we refer to the spectral peak frequency $\omega_p$ as characteristic one unless otherwise stated.

Quite often results of wave studies are recapitulated in the form of power-law functions of the dimensionless time duration $\tau=tg/U_h$ or the fetch $\chi=xg/U_h^2$, as follows
\begin{subeqnarray}\label{eq:exp_law}
  \varepsilon = \varepsilon_0 \tau^{p_{\tau}},\qquad \sigma = \sigma_0 \tau^{-q_\tau};\\
  \varepsilon = \varepsilon_0 \chi^{p_{\chi}},\qquad \sigma = \sigma_0 \chi^{-q_\chi},
\end{subeqnarray}
that already implies, in a manner, universality of wind wave growth.
However, the coefficients $\varepsilon_0,\,\sigma_0$ and the exponents $p_\tau\,(p_\chi),\,q_\tau\,(q_\chi)$ of such parameterizations vary in a certain moderately wide range \citep[e.g. $0.7 < p_\chi < 1.1$, see Table~2][]{BBRZ2007}. Evidence of this variability sometimes leads to the scepsis about existence of any universality in the wind-driven seas:\\
\emph{`Perhaps it is time to abandon the idea that a universal power law for non-dimensional fetch-limited growth rate is anything more than an idealization'. \citep[][p.477]{DonelanEtal1992}.}

We do not share this ultra-radical viewpoint. We consider that the very fact of power-like dependence of energy and mean frequency on fetch and duration is significant and is in need of a theoretical explanation. A number of reasons can be called to explain the lack of complete universality: the  inadequacy of power-law fit, the complexity of wind-wave interaction, the irrelevance of scaling by a mean wind speed with no account of the air flow stratification, gustiness \textit{etc}. All these issues imply a leading effect of wind forcing without accounting for all the complexity of wind-sea dynamics. It leaves (intentionally or unintentionally) inherently nonlinear dynamics of wind waves beyond the discussion.

In this paper we come back to the problem of balance between various physical mechanisms governing wind-wave growth. In contrast to the conventional `common-sense' understanding of this well-known natural phenomenon we develop an alternative paradigm of weakly turbulent wind-driven seas where the universality of wind-wave growth is determined, first of all, by features of nonlinear wave-wave interactions in a random field of water waves.

Thus, we show that for conventional duration- and fetch-limited setups wind-wave growth can be presented in a remarkably concise but, somewhat, paradoxical form:
\begin{equation}\label{eq:key_result}
  \mu^4 \nu =\alpha_0^3.
\end{equation}
Here $\alpha_0 \approx 0.7$ is a universal constant while  $\mu$, the wave steepness, is given by:
\begin{equation}\label{def:steepness}
  \mu=\frac{E^{1/2}\omega_p^2}{g}.
\end{equation}
The `number of waves' $\nu$ in a spatially homogeneous ocean (duration-limited growth) is defined as follows
\begin{equation}\label{eq:nu}
  \nu=\omega_p t.
\end{equation}
For spatial wave growth (fetch-limited case) the coefficient of proportionality $C_f$ that appears in the equivalent expression $\nu=C_f |\kv_p|x$ ($\kv_p$ -- wavevector of spectral peak) is close to the ratio between the  phase and group velocities $C_{ph}/C_{g}=2$. The universal constant $\alpha_0$ in (\ref{eq:key_result}) is analogue of the Kolmogorov-Zakharov constants of wave turbulence theory \citep[e.g.][]{ZakhFalLvov92}.

We believe that this universality is explained by two factors:
\begin{itemize}
  \item Dominance of nonlinear interactions in the balance of energy in wind-driven seas;
  \item Ubiquity of self-similar regimes in wind-driven seas \citep[e.g.][]{Zakharov2005NPG,BPRZ2005,BBRZ2007,GBB2011}.
\end{itemize}
The relationship (\ref{eq:key_result})  presented as a universal law of growth of wind waves looks paradoxical. At the very first glance, wave growth driven by wind does not depend on characteristics of the wind !!! In fact, the effect of wind is recorded by inherent parameters of wave field: wave steepness $\mu$ and dimensionless time -- number of instant wave periods $\omega_p t$ ($|\kv_p| x$). From the viewpoint of the theory of wave turbulence the surprising fact is not the absence itself of wind parameters in (\ref{eq:key_result}) but the extreme conciseness and the physical transparency of this universal relationship.

The theory  of self-similar wind-driven seas developed in \citet{Zakharov2005NPG,BPRZ2005,BBRZ2007} showed definitely that all the complexity of wave growth can be documented in a form that does not contain parameters of wind-wave coupling explicitly. The  nonlinear dynamics of water waves provides a balance of instant wave energy and net (result of generation and dissipation) flux of energy to waves quite similar to the balance in strong hydrodynamical turbulence. This dynamics is not sensitive to the details of wind-wave coupling but is determined by integral properties of this coupling. The power-like fits given by (\ref{eq:exp_law}) have been shown to be self-similar solutions of the model that operates with total energy and total net wave input only. For these solutions the inherent features of nonlinear wave dynamics can be expressed in remarkably elegant form of simple algebraic links  between the exponents $p_\tau\, (p_\chi),\, q_\tau\, (q_\chi)$ and the pre-exponents $\varepsilon_0,\, \sigma_0$ in (\ref{eq:exp_law}) \citep{Zakharov2005NPG,BBRZ2007}. Thus, one has two families (duration- and fetch-limited cases) of self-similar solutions indexed by two independent parameters, say, $p_\tau\,(p_\chi)$ and $\varepsilon_0$. The extensive analysis of numerical \citep{BBZR2008} and experimental results \citep{Zakharov2005NPG,BBRZ2007,BBPRZ2007Honolulu} showed the relevance of these families of self-similar solutions and experimental power-law fits (\ref{eq:exp_law}) to the problem of wind-wave growth. One should stress here, that this analysis was focused on partial solutions, i.e. on solutions for particular indexes $p_\tau\,(p_\chi)$ and, thus, was rigidly linked to a power-law dependency.

A prospect to develop an adiabatic approximation for these families of  self-similar solutions have been sketched by \citet{BBRZ2007} but not detailed properly. In the present paper we revisit this problem. We treat the invariant (\ref{eq:key_result}) as an adiabatic one that allows one for switching between self-similar solutions corresponding to different parameters $p_\tau\, (p_\chi)$ slowly varying in time or space. Thus, the compact law of wind wave growth in the form (\ref{eq:key_result}) manifests a number of remarkable features. It
 \begin{itemize}
  \item does not contain explicitly parameters of wind-wave coupling when being a \emph{`law of wind wave growth'};
  \item does not contain explicitly parameters of adiabaticity (e.g. the exponent $p_\tau\,(p_\chi)$ as a slowly varying parameter) when
 being an \emph{`adiabatic approximation'} for the problem discussed;
 \item does not refer to the initial state of wave field. This property reflects an inherently nonlinear nature of the phenomenon when the initial state appears to be completely forgotten.
\end{itemize}

The paper is aimed at presenting the above paradoxes and their treatment both in terms of mathematical consideration and experimental evidences of the physical law (\ref{eq:key_result}).

We start with a brief theoretical overview, more details being found in paper series \citep[e.g.][]{BPRZ2002,PRZ2003,Zakharov2005NPG,BPRZ2005,BBRZ2007,BBZR2008, Zakharov2010Scr}. We describe the families of the self-similar solutions as a specific alphabet of wind-wave growth (S.~Badulin `ABC of wind-wave growth', XVII conference Waves in Shallow Water Environment, Brest, France, 2010, \verb"http://wave.ocean.ru/badulin/ABC_WISE2010.ppt") where the indexes of the solutions are associated with physically important reference cases of wave growth.

Then we present the key result of this work, the invariant (\ref{eq:key_result}), that can be explained by the following way. In a typical situation the spectra of the wind-driven waves are described by self-similar solutions of the Hasselmann kinetic equation. There are two families of such solutions: one for the fetch-limited, the second one for duration-limited regime. Both depend on two free parameters which are defined by wind-wave interaction and vary slowly on time or fetch. However, relation (\ref{eq:key_result}) does not depend on values of these parameters. This explains universality of (\ref{eq:key_result}). This is remarkable  that the values of the universality parameter $\alpha_0$ are almost the same in both cases of duration- and fetch-limited growth. Self-similarity explains  power-law dependencies (\ref{eq:exp_law}) this universality results from.
  These facts open fair prospects to the analysis of numerical and experimental results.

A number of theoretico-empirical models of wave growth are based essentially on physical scale of wind speed and power-law dependencies of dimensionless wave height on wave period. These models and their reference exponents are well known as the \citet{Toba1972} law of $3/2$, \citet{Hass_ross_muller_sell76} law of $5/3$ and \citet{ZakhZasl83b} case of $4/3$. We show that the physical scales of time duration or fetch are able to replace the conventional wind speed scaling fairly well. The corresponding dependencies within the new scaling give two different exponents $5/2$ for fetch- and $9/4$ for duration-limited cases. The gain of the new scaling is two-fold. First, it eliminates any questions on features of wind wave coupling when the mean wind speed alone cannot reflect the complexity of this coupling in its full. Secondly, the duration- and fetch-limited cases can be discriminated based on the simple difference of the corresponding exponents. We, thus, show that the `unusual' exponents $5/2$ and $9/4$ are consistent with the previous results fairly well: the parametric theoretico-empirical model by \citet{Hass_ross_muller_sell76} can be reformulated easily by using  the new scaling. Our reminiscence shows that the empirical parameterizations used extensively in the latter work can be avoided easily within our purely theoretical approach by accepting the following key assumptions:
\begin{itemize}
      \item the leading role of nonlinear transfer;
      \item the quasi-universality of wave spectrum shapes \citep[\textit{shape invariance} in words by][]{Hass_ross_muller_sell76};
    \item the adiabaticity of wave growth, i.e. slow variations of the parameters of  self-similar solutions in fetch or time.
\end{itemize}

The simple relationship (\ref{eq:key_result}) finds its perfect justification in a number of examples presented in this study.  All the data for the verification of the theoretical result have been obtained previously with no reference to this work. We revisit a collection of experimental results of field studies \citep[see][and refs. therein]{Hwang2006,Hwangetal2011}, simulations by \citet{BPRZ2002,BPRZ2005,BBRZ2007,BBZR2008,ZRPB2012} and wind-wave tank experiments by \citet{Toba1972,CaulliezEtAl2008, BadulinCaulliez2009}. A historical tour to the brilliant work by \citet{SverdrupMunk1947} turns us back to the concept of significant wave height as an effective alternative to the classical spectral description of wind seas.

Future studies will provide undoubtedly new evidences of correctness of the physical law (\ref{eq:key_result}) and, more generally, will show the adequacy of the paradigm of weak turbulence for describing  wind-driven seas.

\section{Invariant form of the self-similar solutions for growing wind seas}

In this work we follow statistical description of a random field of weakly nonlinear wind-driven  waves under effectы of wind forcing and wave dissipation. Within this approach the spectral density of the wave action $N(\kv,\xv,t)$ as function of wavenumber $\kv$, spatial coordinate $\xv=(x,y)$ and time $t$ can be described by the kinetic equation \citep{Hass_62} as follows
\begin{equation}
\label{eq:Kin_full} \frac{\partial N_{\kv}}{\partial t} +
\nabla_{\kv} \omega_{\kv} {\nabla_\xv N_{\kv}} = S_{nl}\left[ N(
\kv)\right]+S_{in}+S_{diss}.
\end{equation}
The idea of balance between wind input $S_{in}$, wave dissipation $S_{diss}$ and wave-wave interactions  $S_{nl}$ has been circulated long before the World War II \citep[e.g.][]{SverdrupMunk1947,Lavrenov98_book}. The start of the modern concept of spectral balance of wind-wave field is usually attributed to the paper by \citet{Gelci1957} where all the terms in  (\ref{eq:Kin_full}) have been treated as wave scale dependent.

The milestone papers of early sixties by Klauss \citet{Hass_62,Hass_63b,Hass_63a} provided a consistent physical description of the term of four-wave resonant interactions $S_{nl}$. The role of these interactions in the evolution of wind-driven waves has been recognized but has not been realized in its full. Key  results of the theory of wave turbulence \citep{ZakhFil66,Katz_Kontor71,Katz_Kontor74,Katz_Kontor75,ZakhZasl83b,ZakhFalLvov92} for the kinetic equation (\ref{eq:Kin_full}) remained beyond the  chief topics of the wind-wave community. Today, the terms of wind input and  wave dissipation are considered as the key ones in the kinetic equation (\ref{eq:Kin_full}) and their study continues to attract  most  efforts of researchers.

The knowledge today of both terms $S_{in}$ and $S_{diss}$ in the right-hand side of (\ref{eq:Kin_full}) is based mostly on empirical parameterizations. It represents an additional problem for wind-wave studies when correct modelling of wave evolution requires tuning to the  features of a particular environment.

In this paper we present results that do not depend on these features and, moreover, do not contain any parameters associated with wave generation or dissipation phenomena explicitly. The physical roots of this surprising result is in the  leading role of the wave-wave interaction term $S_{nl}$ \citep[e.g.][]{Hass_etal73,YoungVledder93,BPRZ2005,ZakharovBadulin2011DAN}: the effects of external forcing appear to be fairly well documented by the intrinsic parameters of nonlinear wave field.

Following the previous works \citep[e.g.][]{Zakharov2005NPG,BPRZ2005,BBRZ2007} consider an asymptotic model describing  wind-driven seas. Assuming wave-wave interactions to be dominant as compared to wind forcing and wave dissipation one can split (\ref{eq:Kin_full}) into two equations. In terms of energy spectral density $E(\kv, \xv,t)$ the latter takes the form:
\begin{subeqnarray}
\label{eq:Kin_split}
  \frac{d E_k}{d t} & = & S_{nl};\\
  \frac{d\langle E_k\rangle}{d t} &=& \langle S_{in}+S_{diss}\rangle.
\end{subeqnarray}
The angle brackets in (\ref{eq:Kin_split}) denote integration over wave scales (i.e. in wave vector space). The first equation (\ref{eq:Kin_split}{\it a}) describes the effect of resonant wave-wave interactions only. The second one (\ref{eq:Kin_split}{\it b}) imposes closure conditions corresponding to the  balance of the total energy: net input (input and dissipation) is equal to growth rate of the total wave energy.

A  breakthrough can be made for deep water waves when the wave dispersion relation and the wave-wave interaction term $S_{nl}$ are homogeneous functions of the spectral density $E(\kv,\xv,t)$ and the wave vector $\kv$, i.e.
\begin{equation}
S_{nl}\left[c E(d \kv)\right]=c^3
d^{17/2}S_{nl}\left[ E( \kv)\right] \label{eq:Homo}
\end{equation}
with $c$ and $d$ being positive coefficients.
This important property allows one to look for self-similar solutions as  functions of time (fetch) and wave frequency (wave number).

\subsection{Power-law dependencies in the self-similar solutions}
Now we briefly outline major features of self-similar solutions for the system (\ref{eq:Kin_split}), details are given in Appendix.

Let us introduce dimensionless variables for the model (\ref{eq:Kin_split}) as follows \citep{Zakharov2005NPG,BPRZ2005,BBRZ2007}
\begin{eqnarray}
\chiv = \xv/l_0; \quad & \tilde \kv  = \kv l_0 \nonumber\\
  \tau = t/t_0; \quad &  \tilde \omega = \omega \sqrt{l_0/g}=\sqrt{ |\tilde\kv|} \label{def:adim}\\
    \tilde E(\tilde \kv) = E(\kv) /l_0^4; \quad & \tilde E(\kv)= E/ l_0^2 \nonumber
\end{eqnarray}
Time and length scales $t_0,\,l_0$ are arbitrary in the deep water case.

For the duration-limited setup one has in (\ref{eq:Kin_split})
\[
\frac{d}{dt} \rightarrow \frac{\partial }{\partial t}
\]
and the  solution in the form of the so-called incomplete self-similarity looks like
\begin{equation}\label{eq:SSadimtext}
  \tilde E(\tilde\kv,\tau)= a_\tau \tau^{p_\tau+4q_\tau} \Phi_{p_\tau}(\xiv)
\end{equation}
where $\xiv= b_\tau \kv t^{2q_\tau}$. Substitution of (\ref{eq:SSadimtext}) to (\ref{eq:Kin_split}) leads to two `magic links' of exponents
\begin{equation}\label{eq:p2qdur}
  q_\tau=\frac{2 p_\tau +1}{9},
\end{equation}
and pre-exponents of the solution
\begin{equation}
\label{eq:a2bdur}
  a_\tau=b_\tau^{17/4} .
\end{equation}
These useful relationships confirm empirical power-like laws (\ref{eq:exp_law}{\it a}) (see Appendix for details) with
\begin{equation}\label{eq:preexp_dur}
  \varepsilon_0=a^{9/17}_\tau I_\tau; \qquad \sigma_0=a_\tau^{-2/17} J_\tau I_\tau^{-1}.
\end{equation}
Here $I_\tau,\,J_\tau$ are integral expressions of the shape function $\Phi_{p_\tau}(\xiv)$ in (\ref{eq:SSadimtext}) that do not depend on exponent $p_\tau$ explicitly. After combining relationships (\ref{eq:exp_law},\,\ref{eq:p2qdur},\,\ref{eq:preexp_dur}) in the form of the invariant (\ref{eq:key_result}) we observe a remarkable fact: the result  does not depend on time and initial state (pre-exponent $a_\tau$). Moreover, its implicit dependence on  exponent  $p_\tau$ is expressed by integrals of shape function $\Phi_{p_\tau}(\xiv)$. Assuming  \emph{spectral shape invariance}, i.e. integrals $I_\tau,\,J_\tau$ to be constants we get immediately $\alpha_0$ in (\ref{eq:key_result}) is constant. This is what we call \emph{universality of wind-driven seas}.

Note, that the assumption of the spectral shape invariance is introduced here for integral quantities $I_\tau,\,J_\tau$ and does not equivalent to point-by-point matching the shape functions $\Phi_{p_\tau}(\xiv)$ for different exponents $p_\tau$. This assumption has been carefully checked in previous extensive numerical studies \citep{BPRZ2005,BBRZ2007,BBZR2008}. Similar assumption has been exploited by \cite{Hass_ross_muller_sell76}. A brief discussion of this paper will be given below.

The same universality holds in the fetch-limited setup. Now
\[
\frac{d}{dt} \rightarrow \frac{\partial \omega}{\partial k}\frac{\partial}{\partial x}
\]
and the self-similar solution is given by expression
\begin{equation}\label{eq:SSadimfetch}
  \tilde E(\tilde\kv,\chi)= a_\chi \chi^{p_\chi+4q_\chi} \Phi_{p_\chi}(\zetav)
\end{equation}
with $\zetav=b_\chi {\tilde \kv} {\tilde x}^{2q_\chi}$. Again, substitution to (\ref{eq:Kin_split}) gives `magic links' for exponents
\begin{equation}\label{eq:p2qfetch}
  q_\chi=\frac{2 p_\chi +1}{10}
\end{equation}
and pre-exponents
\begin{equation}\label{a2bfetch}
  a_\chi=b_\chi^{9/2}.
\end{equation}
Similarly to the duration-limited case pre-exponents of wave growth in (\ref{eq:exp_law}) are
\begin{equation}\label{eq:preexp_fetch}
  \varepsilon_0=a^{5/9}_\chi I_\chi; \qquad \sigma_0=a_\chi^{-1/9} J_\chi I_\chi^{-1}.
\end{equation}
It is easy to check that the invariant (\ref{eq:key_result}) keeps the same form as one in the duration-limited case with number of waves $\nu$ defined in terms of spatial wave periods. Again, the invariant does not depend on time and initial state (pre-exponent $a_\chi$). Implicit dependence on exponent $p_\chi$ is expressed by shape functions $\Phi_{p_\chi}(\zetav)$ and, again, is weak, as shown by extensive simulations.

\subsection{Universal constant $\alpha_0$, spectral shape invariance and `magic links' of  the self-similar solutions}
\label{sect:shapeinv}
The value of the second `magic link' (\ref{eq:a2bdur},\ref{a2bfetch}) that provides independence of the shape spectrum function $\Phi_{p_\tau}(\xiv)$ ($\Phi_{p_\chi}(\zetav)$) on pre-exponent $a_\tau$ ($a_\chi$) has been  first realized by \citet{BBRZ2007} in the form of the so-called \textit{weakly turbulent law of wind-wave growth} \citep[see eq.~1.9][]{BBRZ2007}
\begin{equation}\label{eq:WTGR}
   \frac{E\omega_p^4}{g^2}=\alpha_{ss}\left(\frac{\omega_p dE/dt}{g^2}\right)^{1/3}.
 \end{equation}
Here $\alpha_{ss}$ is a parameter that depends on the exponent of wave energy growth $p_\tau$ ($p_\chi$) only. This law is useful for a particular solution with a fixed parameter $p_\tau\, (p_\chi)$. It allows for conversion instant wave energy and frequency into instant wave input or, oppositely, for retrieval instant wave field characteristics from temporal or spatial variability of the field \citep[e.g.][]{Badulin2014}.

 The idea to develop an adiabatic approach for (\ref{eq:WTGR}) with  $\alpha_{ss}$ that depends on parameter $p_\tau\, (p_\chi)$ of a family of self-similar solutions looks quite tempting one \citep{BBRZ2007}: one can try to construct an adiabatic solution that depends on slowly varying parameter $p_\tau\,(p_\chi)$.

 A trivial observation gives an elegant way to realize this idea.
Note, that power-law combinations of total energy, frequency and time (fetch) for duration-limited
\begin{equation}
\label{eq:combydur}
 E^2 \omega_p^9 t
\end{equation}
  and for fetch-limited cases
 \begin{equation}
 \label{eq:combyfetch}
 E^2 \omega_p^{10} x
\end{equation}
cancel both an explicit dependence on time (fetch) and on the scaling parameter $a_\tau$ ($a_\chi$). These combinations obey the same homogeneity properties as \textit{weakly turbulent law of wind-wave growth} (\ref{eq:WTGR}) but, in contrast to the latter, they do not depend on parameter $p_\tau\,(p_\chi)$ explicitly but on integrals of the shape function $\Phi_\tau\,(\Phi_\chi)$. These integrals depend weakly on parameter $p_\tau\,(p_\chi)$ as numerical and experimental studies show  \citep[e.g.][]{Hass_ross_muller_sell76,BPRZ2005}. With the assumption of \emph{quasi-universality of spectral shaping} \citep{BPRZ2005} or \emph{spectral shape invariance} \citep{Hass_ross_muller_sell76} the combinations (\ref{eq:combydur},\ref{eq:combyfetch}) give invariants of the families of the self-similar solutions in remarkably simple and concise forms:
\begin{equation}
\label{eq:mu4nuf}
\mu^4 \nu = \alpha_{0(d)}^3 \qquad \textrm{or}  \qquad \mu^4 \nu = \alpha_{0(f)}^3
\end{equation}
with constant $\alpha_{0(d)},\,\alpha_{0(f)}$ -- counterparts of the self-similarity parameter $\alpha_{ss}$ in (\ref{eq:WTGR}).  A remarkable feature of these solutions in contrast to (\ref{eq:WTGR}) is in their independence on the adiabaticity parameter $p_\tau\,(p_\chi)$.

We distinguish different notations for  constants $\alpha_{0(d)},\,\alpha_{0(f)}$ which are, evidently, different for duration- and fetch-limited cases. Following \citet{BBRZ2007} and \citet{GBB2011} one can propose the following estimates:
\begin{eqnarray}\label{def:alpha0d}
  \alpha_{0(d)}=\alpha_{ss}^{(d)}p_\tau^{1/3} \approx 0.70\\
   \label{def:alpha0f} \alpha_{0(f)} =\alpha_{ss}^{(f)}p_\chi^{1/3}\approx 0.62.
\end{eqnarray}
Difference in magnitudes of $\alpha_{0(d)}$ and $\alpha_{0(f)}$ can be related, first, to the spectral shaping, which is unlikely to be the major effect if we accept the spectra invariance.

The correspondence of the duration- and fetch-limited cases comes directly from the relationships between the  partial derivative in time and the convective one when treating the model (\ref{eq:Kin_split}), i.e.
\[
  \frac{d}{dt} \to \frac{\partial }{\partial t} \to V \frac{\partial}{\partial x}
\]
Velocity  $V$  is associated with a mean one obtained by averaging the wave energy flux over wave scales, i.e.
\[
  \langle C_g(\kv) E(\kv)\rangle = V \langle E(\kv) \rangle
\]
Velocity $V$ of the mean energy transfer differs from this associated with the group velocity of the spectral peak
\[
C_{g}(\omega_p)=0.5 \frac{g}{\omega_p}
\]
we are exploiting in our analysis. The latter allows for simple definition of number of waves $\nu$ in the fetch-limited case
\begin{equation}\label{eq:nu4fetch}
  \nu=2 |\kv_p| x.
\end{equation}
Small (about $10$\%) difference between $\alpha_{0(d)}$ and $\alpha_{0(f)}$ in (\ref{def:alpha0d},\ref{def:alpha0f}) can be treated as a difference between  the characteristic velocity $V$ and the group velocity of the spectral peak component $C_g(\kv_p)$. If we are calling for `perfect universality' of our law, i.e. equivalence of $\alpha_{0(f)}$ and $\alpha_{0(d)}$ we have to take into account this difference between $V$ and $C_g(\omega_p)$ in our definition of number of waves $\nu$ in (\ref{eq:nu4fetch})
\begin{equation}\label{eq:V2Cgroup}
\frac{V}{C_g(\omega_p)}=\left(\frac{\alpha_{0(f)}}{\alpha_{0(d)}}\right)^3 \approx 0.7.
\end{equation}
The characteristic velocity $V$ for wind-wave spectra is approximately $30$\% less than the group velocity  of the spectral peak component $C_g(\omega_p)$ that matches quite well the previous experimental results \citep[e.g.][]{EfimovBabanin1991,HwangWang2004,Hwang2006}.

Further we also use  notation $\alpha_0$ without subscript when this does not lead to confusion.


\subsection{ABC of `magic numbers' of wind wave growth }
\label{sect:ABC}
Special values of the exponents $p_\tau\,(p_\chi)$ are of particular interest. They correspond to constant fluxes of momentum, energy or action from air flow to waves. These cases can be presented as a reference set of `magic numbers'  or a peculiar ABC of wind-wave growth (Badulin~S.~I., `ABC of wind wave growth', XVII conference Waves in Shallow Water Environment', Brest, France, 2010, \verb"http://wave.ocean.ru/badulin/ABC_WISE2010.ppt"). Table~\ref{Table1} summarizes this set of exponents \citep[cf.][Table~1]{GBB2011}. All the reference cases, evidently, are linked to different exponents $p_\chi$ and $p_\tau$. At the same time, for all the cases the ratios $p_\tau/(2q_\tau)$ and $p_\chi/(2q_\chi)$ for duration- and fetch-limited setups are identical. This interesting result shows the physical relevance of the ratio $p/(2q)$  and the corresponding single-parametric dependency of wave energy on period. Such dependency can be used for the characterizing features of wind-sea coupling  for the qualitatively different physical scenarios of wave growth associated with constant momentum or energy or wave action flux to waves. An example of such a diagnosis can be found in the recent paper by \citet{BadulinGrigorieva2012}.
\begin{table}
\centering
\begin{tabular}{lc  c cccc cccc}
  \hline
  Case & & $p/(2q) \quad $ & $r_\tau$ \quad & $p_\tau $ \quad & $m_\tau$ \quad & $q_\tau$ \quad & $r_\chi$ \quad &$p_\chi $ \quad & $m_\chi$ \quad & $q_\chi$ \quad \\[7pt]
  \hline \\
  A. & $dM/dt = \textrm{const}$ & $5/3$ & $ 13/7$& $10/7$ & $1$ & $3/7$ & $ 13/10$& $1$ & $7/10$ & $3/10$\\[3pt]
  B. & $dE/dt = \textrm{const}$ & $3/2$ & $4/3$& $1$ & $2/3$& $1/3$ & $ 1$& $3/4$ & $1/2$ & $1/4$\\[3pt]
  C. & $dN/dt = \textrm{const}$& $4/3$ & $1$ & $8/11$ & $5/11$& $3/11$ & $ 11/14$& $4/7$ & $5/14$ & $3/14$\\ [3pt]
  \hline
  \end{tabular}
  \caption{Exponents of self-similar solutions for reference cases of wind-wave growth.  The exponents are given for total wave action --   $r_\tau\,(r_\chi)$, wave energy -- $p_\tau\,(p_\chi)$ and wave momentum -- $m_\tau\,(m_\chi)$ \citep[see for details][]{BBRZ2007}}.
  \label{Table1}
\end{table}

\begin{figure}
  \centering
 \includegraphics[scale=0.5,angle=-90]{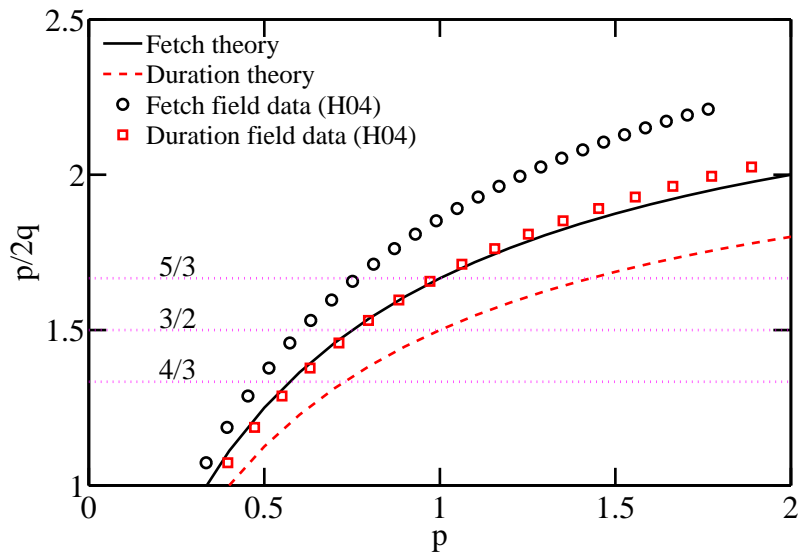}
  \caption{Stages of wind-wave growth in terms of exponents of self-similar solutions (\ref{eq:SSadimtext}, \ref{eq:SSadimfetch}). Wave growth corresponds to evolution from upper right to the lower left corner of the sketch. Symbols correspond to power-like fits of wave growth curves by \citet{HwangWang2004}}\label{fig1}
\end{figure}

{Fig.~\ref{fig1} represents the wind-wave evolution in terms of the theoretical paradigm of the self-similar wind-driven seas. As an alternative to the conventional wind speed scaling \citep[e.g.][fig.~10]{GBB2011} and the corresponding non-dimensional parameter of wave age (ratio of a characteristic wave phase speed to wind velocity) one can consider wave development as a gradual transition from high to lower exponents $p_\tau\,(p_\chi)$, i.e. as a travel from the upper right to the lower left of fig.~\ref{fig1}. The successive transition over the three reference values $A,\,B,\,C$ in this figure expresses the evolution of wind-wave coupling from a state of relatively fast growth at a permanent wave momentum production (case $A$) to a slowly growing under permanent wave action flux pre-mature sea  (case $C$). The theory  shows evolution similar to those of field data but slightly underestimates the corresponding $p/(2q)$ values \citep[see data by][given by symbols in fig.~\ref{fig1}]{HwangWang2004}.
}

Note, that the single-parameter representation of wave growth is well-known, e.g. the famous \citet{Toba1972} law of $3/2$ or the parameterization proposed by \citet{Hass_ross_muller_sell76} (law of $5/3$).
Our universality law (\ref{eq:key_result}) can be considered as an adiabatic invariant of the model (\ref{eq:Kin_split}) valid when parameters of the self-similar solutions $p_\tau\,(p_\chi)$ (or $p_\tau/(2q_\tau)$ and $p_\chi/(2q_\chi)$ of the single-parametric dependencies) are evolving slowly enough with time or fetch. A remarkable feature of (\ref{eq:key_result}) in contrast to other physical systems is that the `magic links' (\ref{eq:p2qdur}, \ref{eq:p2qfetch}, \ref{eq:a2bdur}, \ref{a2bfetch}) make the invariant independent, first, on the adiabaticity parameter ($p_\tau\,(p_\chi)$), and, what is more unusual, on the initial state of the system. The latter reflects the property of the inherently nonlinear system to `forget' effectively its initial state. This is why we can regard the invariant (\ref{eq:key_result}) as one expressing the property of universality of wind wave growth.

\section{Physical scaling of growing wind seas and the first test of the universality of wind wave growth}
{The theoretical results of the previous section look paradoxical and contradictory to a common sense understanding of wind wave dynamics: the invariant (\ref{eq:key_result}) does not refer to any wind parameters. The effect of wind with all the complexity of wind wave coupling appears to be embedded  into intrinsic wave parameters -- wave steepness $\mu$ and dimensionless wave lifetime $\nu$. Thus, the common sense formula \emph{`wind rules waves'} should be replaced by a new one. }

{\emph{Waves keep a wind chronicle} as a balance of their lifetime  $\nu$ and wave steepness $\mu$ (i.e. nonlinearity). The new formula, first, implies a new physical scaling that will be introduced in this section and used further for analysis of experimental and numerical results. Secondly, the consistency of the new formulation with previous experimental and theoretical results will be detailed in order to show the correspondence of this formulation with results inherently based on a wind speed scaling. We will show that this wind speed scaling is, in a sense, excessive: elimination of wind speed from the well-known parameterizations of wind wave growth leads exactly to the wind free relationships proposed here (e.g. eq.~\ref{eq:key_result}).
}

\subsection{Physical scaling of self-similar wave growth}
The invariant (\ref{eq:key_result}) of self-similar solutions for wind driven seas can be written in a form of dependency of wave height on wave period. An example of such dependency is the famous \citet{Toba1972} law of $3/2$ (the dimensionless wave height  is proportional to the power $3/2$ of non-dimensional wave period). The key difference of the new dependency lies, first,  in the  physical scaling. The conventional scaling is based on wind speed (\ref{eq:EnND}, \ref{eq:FreqND}) while the new scaling implied by invariant (\ref{eq:key_result}) is \emph{wind speed free}, containing no wind parameters but fetch or time of wave development.

Let us introduce the dimensionless wave height and period for the fetch-limited case as follows:
\begin{equation}\label{def:HTndFetch}
 \tilde H = \frac{H_s}{x}; \qquad \tilde T=T\sqrt{\frac{g}{8\upi^2 x}}
\end{equation}
for fetch $x$, the spectral peak period $T$ and the significant wave height $H_s=4\sqrt{E}$. For the duration-limited case similar quantities can be introduced as:
\begin{equation}\label{def:HTndDur}
 \tilde H = \frac{H_s}{g t^2}; \qquad \tilde T=\frac{T}{2\upi t}.
\end{equation}
The dimensionless periods defined by (\ref{def:HTndFetch},\ref{def:HTndDur}) have a simple physical meaning as they express wave lifetime in terms of number of instant  temporal or spatial wave periods. For the duration-limited case (\ref{def:HTndDur}) it follows:
\begin{equation}\label{eq:T2nu}
  \tilde T=\nu^{-1}
\end{equation}
and for the  fetch-limited case (\ref{def:HTndFetch})
\begin{equation}\label{eq:T2nu12}
  \tilde T=\nu^{-1/2}.
\end{equation}
Definitions (\ref{eq:T2nu},\ref{eq:T2nu12}) represent \emph{a kinematic treatment of invariant} (\ref{eq:key_result}): the instant wave steepness is thus determined by time (distance) of wave evolution expressed in  dimensionless instant wave period.

One can propose \emph{ a dynamical interpretation} of (\ref{eq:key_result}). Note that $\mu^4$ defined by (\ref{def:steepness}) gives a scale of nonlinear relaxation of weakly nonlinear deep water waves. Using recent estimates \citep[see eqs.~22,23 in][]{ZakharovBadulin2011DAN} one can treat time $t$ as a \textit{dynamical lifetime}
\begin{equation}\label{eq:dinage}
t  = B\alpha_0\tau_{nl}
\end{equation}
where
\[
\tau_{nl}= \left(B \omega_p \mu^4\right)^{-1}
\]
is an estimate of the characteristic time of nonlinear relaxation of weakly nonlinear deep water wave field. $B$ is a large coefficient as shown by \citet{ZakharovBadulin2011DAN} ($B=36\pi$ in the limit of narrow angular wave spectrum, $B=22.5\pi$ -- for isotropic wave field). Thus, (\ref{eq:key_result}) says that wave age $t \cdot \tau_{nl}^{-1}$ measured in nonlinear relaxation scale $\tau_{nl}$ remains constant for growing wind waves. In accordance with (\ref{eq:dinage}) and estimates by \citet{ZakharovBadulin2011DAN} this wave age does not exceed one hundred relaxation times $\tau_{nl}$.

\subsection{Single-parameter approximation by \citet{Hass_ross_muller_sell76} and power-law dependencies of wave growth}
Invariant  (\ref{eq:key_result}) being presented in terms of dependence of dimensionless height $\tilde H$ on dimensionless period $\tilde T$ leads to somewhat confusing results. For duration-limited growth one gets `the law of $9/4$' written as
\begin{equation}\label{eq:H2TDur}
  \tilde H = 4 \alpha_{0(d)}^{3/4} \tilde T^{9/4} \approx 3.06 \tilde T^{9/4}.
\end{equation}
The fetch-limited dependence differs from (\ref{eq:H2TDur}) by a factor of $2$, but, what is more significant, by the exponent. Wave growth in space follows the `$5/2$ law' given by
\begin{equation}\label{eq:H2TFetch}
  \tilde H = 8 \alpha_{0(f)}^{3/4} \tilde T^{5/2} \approx 5.59 \tilde T^{5/2} .
\end{equation}
The difference between exponents in (\ref{eq:H2TDur}) and (\ref{eq:H2TFetch}) provides a quantitative criterium for discriminating  spatial and temporal scenarios of wave growth.

The exponents $9/4$ and $5/2$  in  (\ref{eq:H2TDur},\ref{eq:H2TFetch}) look confusing in view  of their wind-scaled counterparts \citep[e.g.][]{Toba1972,Hass_ross_muller_sell76}. Moreover, the key result itself that the dependence of wave height on period in (\ref{eq:H2TDur},\ref{eq:H2TFetch}) does not contain any reference to wind speed looks strange and contradictory to the today basics of wind-wave physics.  In fact, this result consists with the previous studies perfectly well.

{\citet{Hass_ross_muller_sell76} proposed a parametric wave prediction model based on the solid background of the JONSWAP experiment \citep{Hass_etal73} and an extensive theoretical analysis of the properties of the kinetic equation (\ref{eq:Kin_full}). Many points of this theory follow our paradigm of dominating effect of nonlinear transfer on wave growth. We will fix these points below in separate subsections and detail them  in the context of our approach.}

\subsubsection{Self-similarity of the spectral shape}
\citet{Hass_ross_muller_sell76} started with the JONSWAP parameterization of wave spectrum as а  function of five parameters \citep{Hass_etal73}. This step corresponds to the first equation of our model (\ref{eq:Kin_split}{\emph a}) that prescribes a self-similar shape function $\Phi_{p_\tau}(\xiv)\,(\Phi_{p_{\chi}}(\zetav))$ in  (\ref{eq:SSadimtext},\ref{eq:SSadimfetch}). The function $\Phi_{p_\tau}(\xiv)$ (or its counterpart for the fetch-limited setup) can be found as a solution of an integro-differential equation  \citep[see][eq.~91 and eq.~\ref{eq:ss_detail} of Appendix]{BPRZ2005} while parameters of this solution, exponents and pre-exponents, are dictated by the second equation of the model (\ref{eq:Kin_split}{\emph b}) -- condition of the integral balance of wave energy. `The magic links' (\ref{eq:p2qdur},\ref{eq:p2qfetch}) between the exponents $p_\tau$ and $q_\tau$ ($p_\chi$ and $q_\chi$)  is a stem of our analysis while the particular features of the shape function $\Phi_{p_\tau}\,(\Phi_{p_\chi})$ are of no importance.

In contrast to our theoretical approach \citet{Hass_ross_muller_sell76} exploit  empirical links between the five parameters of the  JONSWAP spectrum and wind speed. All these links can be regarded as an assumption of self-similarity of the JONSWAP  wind-wave spectrum that allows for an effective parameterization of the effects of both nonlinear wave-wave interactions and quasi-linear terms of wind input and wave dissipation. The resulting set of partial differential equations describes dependence of the JONSWAP spectrum parameters on wind speed and, then, is used for closing the balance of wind-wave energy.

\subsubsection{Balance between nonlinear transfer and external forcing}
A condition of permanent wind stress  exerted to waves (in other words, constant wave momentum flux or constant drag coefficient) is introduced to close the balance of the wave energy in the \citet{Hass_ross_muller_sell76} model (see eq.~3.4 herein). It fixes `a magic link' of the model as $q/p=3/10$ for both duration- and fetch-limited cases.

Alternatively, the balance  equation (\ref{eq:Kin_split}{\emph b}) and  its adiabatic counterpart (\ref{eq:WTGR}) treat the balance in terms of total fluxes of energy, momentum or wave action without any explicit reference to wind speed scaling and particular type of wind-wave coupling. As a result, the energy growth  rate $p_\tau\,(p_\chi)$ appears to be linked to the frequency downshift exponent $q_\tau\,(q_\chi)$  exclusively by properties of homogeneity of the kinetic equation (\ref{eq:Homo}), but not by a set of empirical parameters of the JONSWAP spectrum as in the model by \citet{Hass_ross_muller_sell76}. Within our approach the dependencies of spectral fluxes on time or fetch are not restricted by  additional assumptions of constant wind stress or any other specific scenarios of wave input. Thus, our approach can be regarded as more general than the theoretico-empirical model by \citet{Hass_ross_muller_sell76}.

\subsubsection{Quasi-universality of wind wave spectra}
Another parallel between both theories can be found in the assumption of \emph{quasi-universality} \citep[as defined by][see sect.~\ref{sect:shapeinv} of this paper]{BBRZ2007} or \emph{shape invariance of wave spectra} \citep[in words of][]{Hass_ross_muller_sell76}.  Under this assumption final expressions for self-similar solutions take quite transparent forms in both theories. Stress, in both theories the quasi-universality (the spectral shape invariance) refers to integral quantities and does not require point-by-point proximity of spectral distributions.

\subsubsection{Self-similar solutions as counterparts of empirical parameterizations}
Self-similar solutions of both approaches, the \citet{Hass_ross_muller_sell76} one and ours follow the conventional empirical fits of wind-wave growth expressed by (\ref{eq:exp_law}\textit{a,b}). In contrast to the empirical treatment both theories predict `magic link' between energy growth exponent and frequency downshift. Our theory gives different linear relationships (\ref{eq:p2qdur}) and (\ref{eq:p2qfetch}) for duration- and fetch-limited cases correspondingly and universal values of the ratios $p_\chi/(2q_\chi)$ and $ p_\tau/(2q_\tau)$ that  depend exclusively on the type of wind wave coupling (the ABC of wind wave growth, see Table~\ref{Table1} and fig.~\ref{fig1}) rather than on a particular setup.

\begin{figure}
  \centering
 \includegraphics[scale=0.5,angle=0]{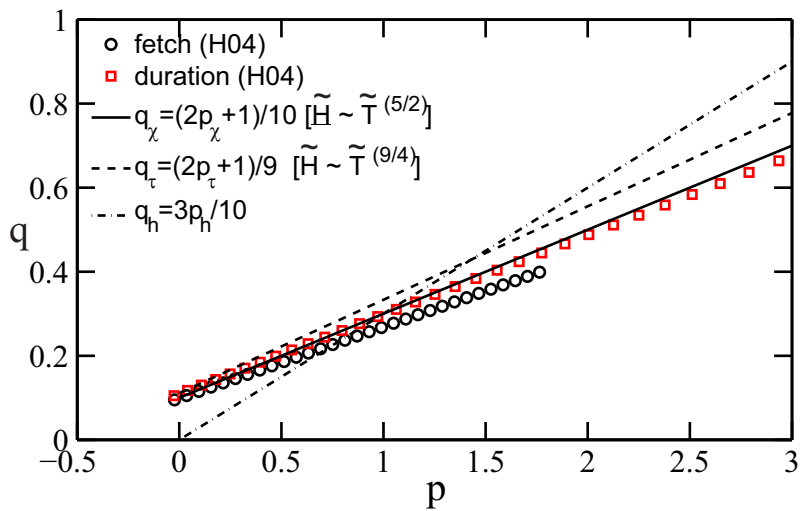}
  \caption{Dependencies of frequency downshift exponent $q$ on energy growth one $p$ for duration- and fetch-limited setups for experimental fits by \citet{HwangWang2004} (symbols), theory of this paper (solid and dashed lines)  and one of \citet{Hass_ross_muller_sell76} (dash-dotted line).}\label{fig2}
\end{figure}

In contrast to our approach \citet{Hass_ross_muller_sell76}  propose the unique dependence for both setups written as follows:
\begin{equation}\label{eq:p2qHass}
  q_{h}=\frac{3}{10} p_{h}
\end{equation}
that fixes the type of wind wave coupling as one corresponding to a permanent wind stress exerted on waves. Thus, the theory by \citet{Hass_ross_muller_sell76} is not consistent with the ABC of wind wave growth presented in sect.~\ref{sect:ABC}, and so, it does not represent the wave field evolution as a transition between the different scenarios of wind-wave coupling. Our theoretical approach and experimental fits by \citet{HwangWang2004} appear to be more general in reflecting diversity of mechanisms of wind wave spectrum evolution. The corresponding dependencies are shown in fig.~\ref{fig2}.

In  sect.~2 we showed how the family of self-similar solutions  can be written in a universal form that does not contain the dependence on exponents $p_\tau\,(p_\chi)$. A remarkable feature of the \citet{Hass_ross_muller_sell76} work is that their solutions can be written in a similar way. Excluding dependence on wind speed one can thus express these solutions in the form (\ref{eq:H2TDur},\ref{eq:H2TFetch}). Following notations of \citet[][eqs.~5.3--5.10 therein]{Hass_ross_muller_sell76} we obtain for duration-
\begin{equation}\label{eq:H2TdurH}
  \tilde H = 4(2\pi)^{9/4}C^{1/2}A^{7/12} {\tilde T}^{9/4}
\end{equation}
and fetch-limited cases:
\begin{equation}\label{eq:H2TfetchH}
  \tilde H = 4(8\pi^2)^{5/4}C^{1/2}A^{5/6} {\tilde T}^{5/2}.
\end{equation}
Here  the parameter of spectral shape invariance can be assumed to be constant $C=5.1\times 10^{-6}$ \citep{Hass_ross_muller_sell76}. The coefficient $A$ in (\ref{eq:H2TdurH}, \ref{eq:H2TfetchH}) depends weakly on exponents of power-like dependence of instant wind speed on fetch or time and can set to be constant as well, i.e. $A=16.8$ for duration- and $A=2.84$ for fetch-limited cases \citep[see][eqs.~5.7--5.10]{Hass_ross_muller_sell76}. Substituting these values into (\ref{eq:H2TdurH},\ref{eq:H2TfetchH}) one gets respectively
\begin{subeqnarray}
  \tilde H & \approx & 2.93 \cdot{\tilde T}^{9/4} \label{eq:HassDur}\\
  \tilde H & \approx & 5.07 \cdot {\tilde T}^{5/2} \label{eq:HassFetch}
\end{subeqnarray}
The coefficients in (\ref{eq:HassDur}{\it a, b}) appear to be quite close to those of our theory (cf. \ref{eq:H2TDur},\ref{eq:H2TFetch}). In terms of our invariant $\alpha_0$ this leads to:
\[
\alpha_{0(d)}=0.660; \qquad \alpha_{0(f)}=0.545\
\]
i.e. the  values are remarkably close to our theoretical values (\ref{def:alpha0d},\ref{def:alpha0f}). The formulas derived by \citet{Carter1982}  on the basis of the theory by \citet{Hass_ross_muller_sell76} give slightly different estimates of the coefficients  in (\ref{eq:HassDur}\textit{a}), i.e. $2.92$ rather than $2.93$ thus with $\alpha_{0(d)}=0.658$, and in (\ref{eq:HassFetch}\textit{b}) $4.99$ instead of $5.07$ thus with $\alpha_{0(f)}=0.533$.

Finalizing this section one should stress a deep correspondence between both approaches: the theoretico-empirical by \citet{Hass_ross_muller_sell76} and the theoretical one developed in this work. Both approaches lead to the same wind-speed-free dependencies (\ref{eq:H2TDur},\ref{eq:H2TFetch}). Independent estimates of the physical invariants also give remarkably close values of $\alpha_{0(d)}$ and $\alpha_{0(f)}$. This can be considered as a positive validation of our approach.

Below we implement the adiabatic invariant (\ref{eq:key_result}) and dimensionless dependencies of wave height on period (\ref{eq:H2TDur},\ref{eq:H2TFetch}) for describing manifestations of wind wave growth as observed in numerical simulations, in-situ and wind-wave tank experiments.

\section{Simulations of wind wave growth}
Self-similarity features of wind wave growth have been depicted in detail previously  in a number of numerical studies. \citet{BPRZ2002,BPRZ2005,BBRZ2007,BBZR2008,Lavrenov2003,PRZ2003} used the algorithm proposed by \citet{Webb78} and developed by \citet{Tracy82} for simulating duration-limited growth within the kinetic equation (\ref{eq:Kin_full}) with an exact collision integral $S_{nl}$. The properties of self-similarity have been checked in terms of explicit relationships for exponents of wave growth and universality of the spectral shape functions $\Phi_{p_\tau}(\xiv)$.   Specific manifestations of the self-similar wave spectrum behavior have been noticed using the so-called Gaussian Quadrature Method (GQM) in \citet{Lavrenov2002Banff,Lavrenov2003}. Based on the latter approach \citet{GagnaireEtal2010} proposed a quasi-exact method to compute the wave-wave interaction term $S_{nl}$. This method has been successfully  used for modelling the fetch-limited growth with different functions of wind input in a wide range of wind speeds \citep{GBB2011}. More recently, the fetch-limited growth has been simulated by \citet{ZRPB2012} within the approach by \citet{Tracy82} and the exact collision integral formulation. Here we refer to the results presented by \citet{BPRZ2002,BPRZ2005,BBRZ2007,BBZR2008,ZRPB2012} for verifying the above theoretical results both in terms of the invariant (\ref{eq:key_result}) and the single-parametric dependencies of wave height on period (\ref{eq:H2TDur},\ref{eq:H2TFetch}).

\subsection{Duration-limited growth}
\begin{figure}
  \centering
 \includegraphics[scale=0.35]{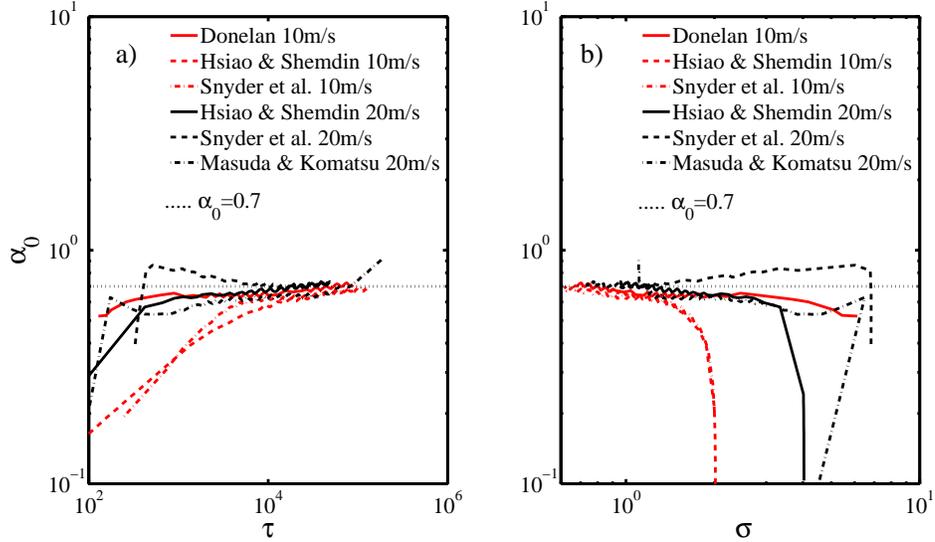}
  \caption{Dependence of  parameter $\alpha_0=(\mu^4 \nu)^{1/3}$ on: (\ref{eq:key_result}) \emph{(a)}  non-dimensional duration $\tau=tg/U_{10}$ and \emph{(b)}  inverse wave age $\sigma=\omega_p U_{10}/g$ in simulations of duration-limited wind wave growth by \citet{BPRZ2002,BPRZ2005,BBPRZ2007Honolulu,BBZR2008}. Simulation setups (wind input parameterization and wind speed) are given in legends. The horizontal dotted line shows theoretical value  $\alpha_{0(d)}=0.7$.
  }\label{fig3}
\end{figure}
Results of simulations by \citet{BPRZ2005} have been used directly for verification of the law (\ref{eq:key_result}) for duration-limited wave growth. Fig.~\ref{fig3} presents somewhat `eclectic' dependencies of the `wind-free' invariant $\alpha_0=(\mu^4 \nu)^{1/3}$ on `wind-scaled' non-dimensional duration $\tau=t g/U_{10}$ (left panel) and inverse wave age $\omega_p U_{10}/g$ (right panel). Wave input functions and wind speed values  used in simulations are shown in legends. The straight line $\alpha_{0(d)}=0.7$ is shown as reference.
All the simulations except the last one \citep[our reproduction of case by][]{Komatsu_Masuda96} have been carried out with a primitive dissipation function associated with a hyper-dissipation at high frequencies \cite[see][for details]{BPRZ2005}. In these cases there is no saturated (fully developed or mature) wind sea state and all the dependencies are tending to  $\alpha_{0(d)}\approx 0.7$ in full agreement with the results of \S~2.  The inverse wave age $\omega_p U_{10}/g$ can reach values less than unity, i.e. waves at the spectral peak can propagate essentially faster than wind speed where conventional models predict no wave growth.

\citet{Komatsu_Masuda96} used a more sophisticated `white-capping' function by \citet{Hass_74} \citep[see also][]{Komen_Hass84} to describe wave dissipation. In this case, wind-sea state is tending to a saturation. This feature is evidenced by a break of the dependency from the general tendency to a saturation at inverse wave age $\omega_p U_{10}/g \gtrsim 1$. Wave steepness $\mu$ and frequency $\omega_p$ are then approaching the limiting values while time $t$ (and, evidently, number of waves $\nu$) continues to grow.

\subsection{Simulations of fetch-limited growth}
\label{sect:FetchSim}
\begin{figure}
  \centering
 \includegraphics[scale=0.35]{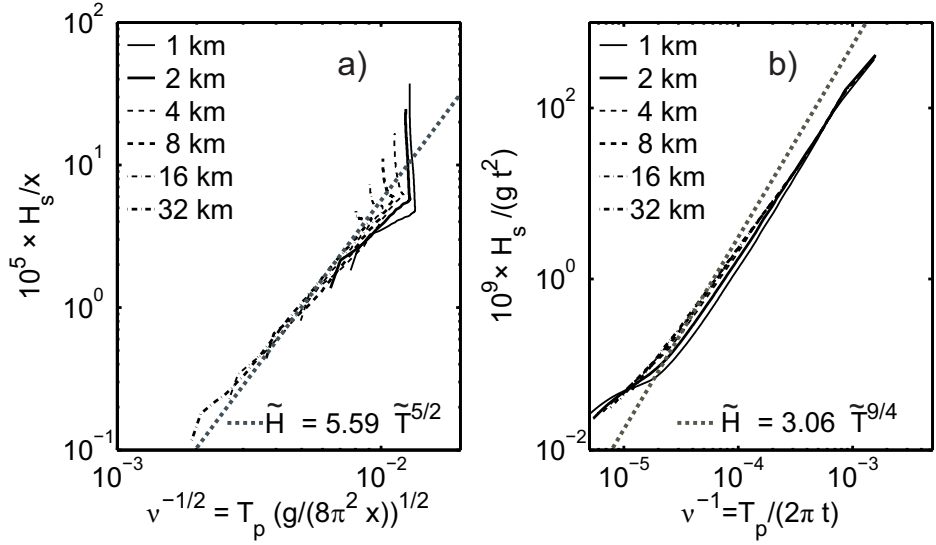}
  \caption{Wave growth curves in simulations of fetch-limited setup  \citep{ZRPB2012} within: \emph{(a)} fetch scaling (\ref{def:HTndFetch}) and \emph{(b)} duration scaling (\ref{def:HTndDur}). Curves are given for fixed fetches $1,\,2,\,4,\,8,\, 16,\, 32$ km (see legends).  Theoretical dependencies (\ref{eq:H2TDur}) are shown by dotted lines. }\label{fig4}
\end{figure}
 The fetch-limited growth  has been simulated by \citet{ZRPB2012} starting from an initial white-noise spatially homogeneous wave field in a coordinate interval 0--41 kilometers  and for times up to $385000$ seconds (approximately 107 hours). Strictly speaking, there is not a classic fetch-limited regime as a stationary state in these simulations. The evolution looks like a sequence of stages where reference dependencies (\ref{eq:H2TDur}, \ref{eq:H2TFetch}) describe intermediate asymptotics. Fig.~\ref{fig4} presents results  by \citet{ZRPB2012} in terms of dependencies on time  at fixed fetches. Curves for log-spaced fetches $1,2,4,8,16$ and $32$ km are shown.

  Fig.~\ref{fig4}{\it a} shows the evolution of wave parameters within the wind free scaling from lower left to upper right corner and demonstrates an impressive coincidence with the $5/2$ power law (\ref{eq:H2TFetch}) over a wide range. Deviations from this dependence are seen at short times (lower left) when wave spectra are far from a self-similar regime. At long times (upper right in fig.~\ref{fig4}\emph{a}) a saturation of the wave field is evidenced by vertical asymptotes of the curves: wave periods are  tending to their finite limits while wave heights are continuing to grow. This effect is  likely associated with a simulation setup  where the  waves are modelled in a coordinate box. At long times when the quasi-stationary state is reaching the upper bound of the box, the quasi-stationary fetch-limited growth then  gives place to a duration-limited scenario.

The time-scaled representation (\ref{def:HTndDur}) in fig.~\ref{fig4}\textit{b} traces the wave evolution in opposite sense when compared with fig.~\ref{fig4}\textit{a}: both $\tilde H$ and $\tilde T$ are decreasing with time. Curves plotted for different fetches appear to be  remarkably close to each other and fit in logarithmic scales the theoretical tangent $9/4$ in a range of more than one decade of $\tilde T$. At longer times (dimensionless periods $\tilde T < 2\cdot 10^{-5}$) the saturation of wave field growth is evidenced by curves deviating well above the reference dependence (\ref{eq:H2TDur}) -- counterparts of the saturation stage in fig.~\ref{fig4}\emph{a}.

Thus, the relationships $\tilde H(\tilde T)$ (\ref{eq:H2TDur}, \ref{eq:H2TFetch}) provide a good reference for the physical analysis of wave evolution. As we show below, the effectiveness of this tool is not limited by the special feature of setups adopted for simulations.

\section{Wind wave growth in field experiments}
In this section we investigate the invariance of wind-wave growth expressed by (\ref{eq:key_result}) for  a collection of field experiments on wind-wave growth. First, we consider a number of experimental  works since the milestone work by \citet{SverdrupMunk1947} and then proceed with in situ data of wave riders of the Field Research Facility \\ (\verb"http://www.frf.usace.army.mil/frf_data.shtml").

\subsection{Classic experiments on duration- and fetch-limited growth}
\begin{figure}
  \centering
 \includegraphics[scale=0.6,angle=-90]{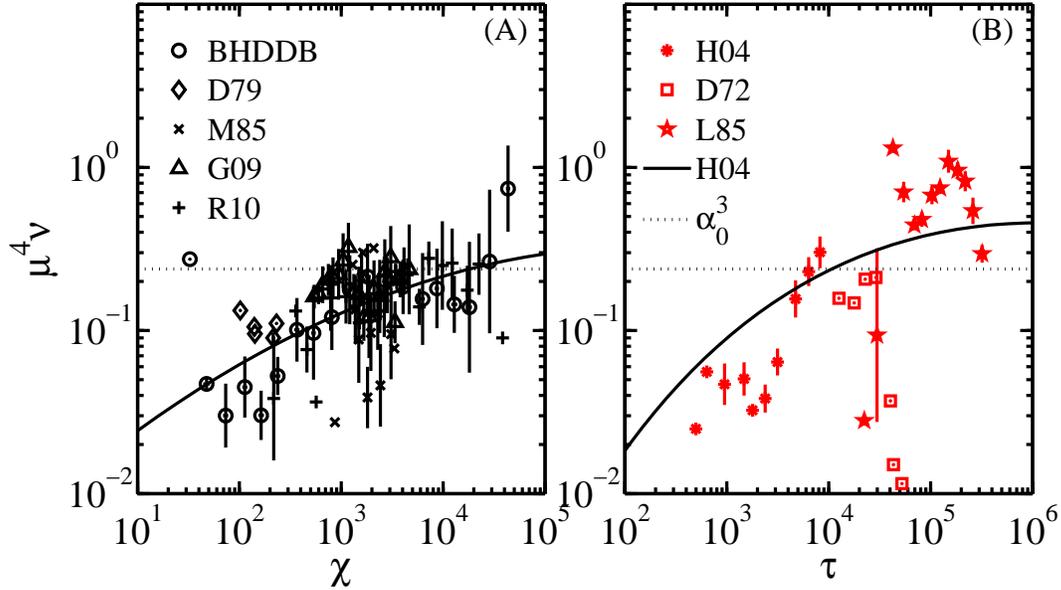}
  \caption{Dependence of wave invariant $\mu^4 \nu $ on: \emph{(A)} non-dimensional fetch $\chi=xg/U_{10}^2$; \emph{(B)} duration $\tau=tg/U_{10}$  scaled by conventional wind speed $U_{10}$ at 10 meters height. Both series of fetch- and duration-limited experiments show a remarkable tendency to the theoretical prediction $\mu^4 \nu=\alpha_0^3 $.}\label{fig5}
\end{figure}

\begin{figure}
  \centering
 \includegraphics[scale=0.6,angle=-90]{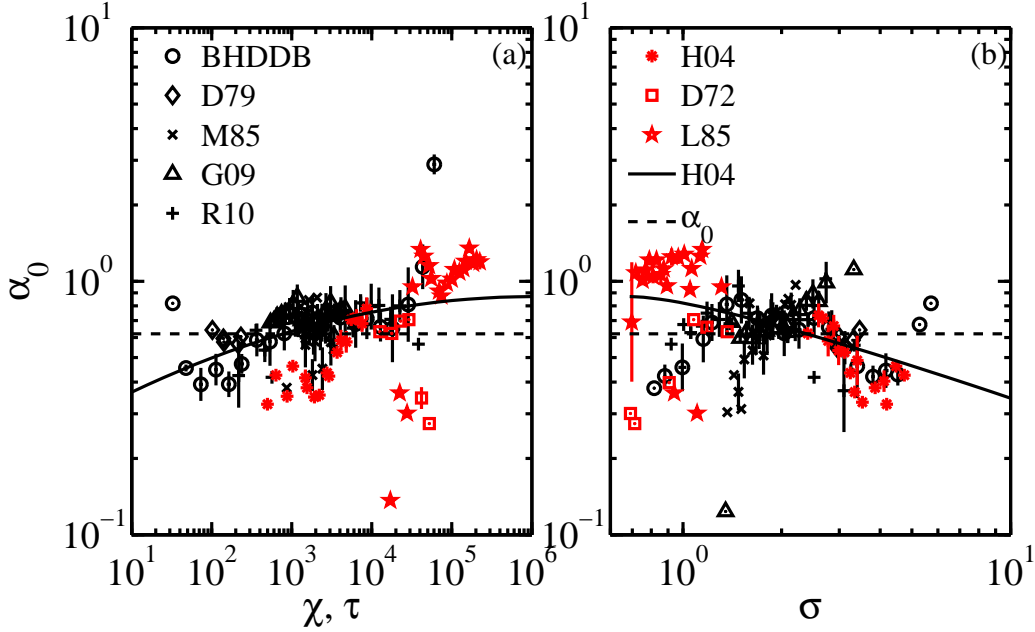}
  \caption{Dependence of experimental estimate of self-similarity parameter $\alpha_0$: \emph{(a)} on non-dimensional fetch $\chi=xg/U_{10}^2$ and duration $\tau=tg/U_{10}$;  \emph{(b)} on non-dimensional frequency of spectral peak $\tilde \omega=\omega_p U_{10}/g$ for combined collection of duration- and fetch-limited data sets. The dashed line shows the theoretical value $\alpha_{0(f)}=0.62$, the solid line is the power-law fit (\ref{eq:empiric_invar1}) for the collection. }\label{fig6}
\end{figure}
As discussed in \citet{HwangWang2004}, two classes of ocean wave measurements are of great importance for the study of generation of ocean waves by wind. The first class is the fetch-limited growth condition, under which the wave development is limited by the available spatial coverage upwind of the measurement location. Over the years, there have been several successful field experiments reported. Extensive reviews and analyses on these datasets have been given \citep[e.g.][]{Kahma_Calkoen92,KahmaCalcoen1994,YoungMon99,HwangWang2004,Hwang2006,BBRZ2007,Hwangetal2011}

The second class is the duration-limited growth condition, under which the wave development is limited by the temporal duration of the steady wind event acting on the water surface. The ideal initial and boundary conditions satisfying the duration-limited wave growth rarely occur in nature and it is no wonder that reports of such data are very scarce. \citet{YoungMon99} presents an extensive review of fetch- and duration-limited wave growth studies. The only duration-limited datasets cited are \citet{SverdrupMunk1947}, \citet{Bretsch1952b,Bretsch1952a} and \citet{Darbyshire1959}, as compiled by \citet{Wiegel1961}. \citet{DeLeonibus} report field data that contain duration growth information. \citet{Liu1985} describes an interesting episode of almost 60 hours quasi-steady wind forcing of wave growth measured by an NDBC wave buoy in Lake Superior. All these data are obtained at later stages of wave development with dimensionless time, $\tau=gt/U_{10}$, greater than about $7000$. By chance, \citet{HwangWang2004} obtained one data set describing the duration growth in the first two hours of wind wave generation and extend the data coverage by more than one order of magnitude ($\tau$ between $498$ and $8801$).

For our quantitative analysis, the fetch or duration data sets require the complete triplets of ($\varepsilon,\,\sigma$ and $\tau$ ) or ($\varepsilon,\,\sigma$ and $\chi$ ). We are able to assemble 9 fetch-limited data sets \citep[][the first 5 data sets are called collectively BHDDB]{Burling1959,Hass_etal73,Donelan1979,MerziGraf1985,Babanin_Soloviev98a,Dobson_al1989,Garciaetal2009,RomeroMelville2010} and 3 duration-limited data sets \citep{DeLeonibus,Liu1985,HwangWang2004} in figs.~\ref{fig5}, \ref{fig6}.

In keeping with the conventional observations that the wave growth functions can be expressed as power-law functions (\ref{eq:exp_law}) but the exponents can vary with the range of dimensionless fetch or duration in different experiments \citet{HwangWang2004}  developed a second order fitting of the power-law functions to the BHDDB fetch-limited data and obtained \citep[cf.\ref{eq:exp_law}, notations are slightly modified as compared with][]{HwangWang2004}:
\begin{subeqnarray}
  \varepsilon &= \varepsilon_{0\chi} \cdot \chi^{p_\chi(\chi)}, \\
  \sigma &= \sigma_{0\chi} \cdot \chi^{-q_\chi(\chi)}
\end{subeqnarray}
where exponents are slowly varying functions of dimensionless fetch $\chi$ given by
\begin{subeqnarray}
  p_\chi (\chi) &=&  \alpha_1+2\alpha_2 \ln\chi \label{eq:pq_Paul_p}\\
  q_\chi (\chi) &=&  \beta_1 + 2\beta_2 \ln \chi \label{eq:pq_Paul_q}
\end{subeqnarray}
with $\varepsilon_{0\chi} =\exp{(-17.6158)}$, $\alpha_1 = 1.7645$, $\alpha_2 = -0.0647$, $\sigma_{0\chi} = \exp{(3.0377)}$, $\beta_1 = 0.3990$, and $\beta_2 = -0.0110$. A sufficient range of the dimensionless fetch $\chi$ for computation is $10^0 - 10^6$.

Note, that slow dependence of exponents $p_\chi,\,q_\chi$ on fetch $\chi$  is in line with the starting point of our theoretical approach: in the course of wave growth the exponents are varying slowly (as logarithms of dimensionless fetch) and are linked to each other by a linear relationship. Excluding fetch $\chi$ in (\ref{eq:pq_Paul_p}\emph{a,b}) one has a counterpart of the `magic link' (cf.~\ref{eq:p2qfetch}) as follows
\begin{equation}\label{eq:p2qPaulFetch}
  q_\chi=\frac{2 P_\chi p_\chi + S_\chi}{10},
\end{equation}
where $P_\chi=0.8501$, $S_\chi=0.9900$  are remarkably  close to the theoretical unity.
The duration-limited growth functions can be derived from a similar approach using a formal conversion of fetch $x$ to time $t$:
\begin{equation}\label{eq:V2Cgr}
 \int_0^t dt=\int_0^x \frac{dx}{C_{gx}}
\end{equation}
where $C_{gx}$ is the downwind component of the wave group velocity. This equation can be expressed in dimensionless form
\[
\tau = \int_0^\chi \frac{ \sigma}{R}\, d \chi,
\]	
 for $\tau=t g/U_{10},\, R=C_{gx}/C_p,\, \chi=xg/U^2_{10}$ and $C_p$ being the phase speed. For a monochromatic wave train, $R=0.5$. For wind seas with wide directional distributions, field measurements by \citet{EfimovBabanin1991} show that $R \approx 0.4$ (cf.~\ref{eq:V2Cgroup}, the theoretical estimate of the ratio). The parameters of duration-limited growth functions can therefore be expressed in terms of those of the fetch-limited growth functions. As $p_\chi(\chi)$ and $q_\chi(\chi)$ are slowly varying functions of dimensionless fetch $\chi$ (because of logarithm in eqs.~\ref{eq:pq_Paul_p}{\it a,b}) one has
\begin{subeqnarray}
  \varepsilon = \varepsilon_{0\tau} \cdot \tau^{p_\tau(\tau)}, \\
  \sigma = \sigma_{0\tau} \cdot \tau^{-q_\tau(\tau)},
\end{subeqnarray}
where
\begin{subeqnarray}
   \varepsilon_{0\tau}  & = & \varepsilon_{0\chi} \left[\frac{R(1-q_\chi)}{\sigma_{0\chi}}\right]^{p_\tau}, \quad \displaystyle{\sigma_{0\tau}  = \sigma_{0\chi}\left[\frac{R(1-q_\chi)}{\sigma_{0\chi}}\right]^{-q_\tau}  } \\
  p_\tau  &= &  \frac{p_\chi}{1-q_\chi},   \qquad  \displaystyle{q_\tau  = \frac{q_\chi}{1-q_\chi}}.
\end{subeqnarray}
Trivial algebra leads to similar relationship between $p_\tau$ and $q_\tau$ (cf.~\ref{eq:p2qPaulFetch}), i.e.
\begin{equation}\label{eq:p2qPaulDur}
   q_\tau=\frac{2P_\tau p_\tau + S_\tau}{9}
\end{equation}
where
\begin{equation}\label{eq:p2qPaulDur1}
  P_\tau=\frac{9P_\chi}{10-S_\chi}, \qquad S_\tau=\frac{9S_\chi}{10-S_\chi}
\end{equation}
and, again, the values of these coefficients
\[
P_\tau=0.8492; \qquad \textrm{and } \qquad S_\tau=0.9889
\]
appear to be quite close to the theoretical unity. Note, that (\ref{eq:p2qPaulDur1}) is consistent with theoretical relationships (\ref{eq:p2qdur}, \ref{eq:p2qfetch}) which derivation does not use explicitly kinematical relationship (\ref{eq:V2Cgr}).

Exploiting the idea of derivation of an adiabatic invariant for growing wind seas (see sect.~2 and eq.~\ref{eq:key_alt}) one can propose an `empirical invariant' for relationships (\ref{eq:p2qPaulFetch}, \ref{eq:p2qPaulDur}) that differs slightly from its theoretical counterpart:
\begin{equation}\label{eq:empiric_invar1}
  (\mu^4 \nu) \chi^{-0.5391+0.0388\ln\chi} = { I_{fetch}}.
\end{equation}
The additional dependence on dimensionless fetch $\chi$ in (\ref{eq:empiric_invar1}) yields a variation by a factor $2.7$ when dimensionless fetch varies over a range of  six orders of magnitude $\chi =10^0-10^6$!!! This difference can be partially explained by the features of these field experiments for which the finite resolution in frequency becomes critical for a correct assessment of the invariant (\ref{eq:key_result}).

Figs.~\ref{fig5} and \ref{fig6} show the results of these field measurements. One can see consistency of the theoretical invariant with experimental data and with  fits (\ref{eq:pq_Paul_p}{\it a,b}) by \citet{HwangWang2004}. Typically, the duration-limited data show larger data scatter reflecting the departures from ideal duration-limited wave generation conditions.

\subsection{Data by \citet{SverdrupMunk1947}}
\begin{figure}
  \centering
   \includegraphics[scale=0.8]{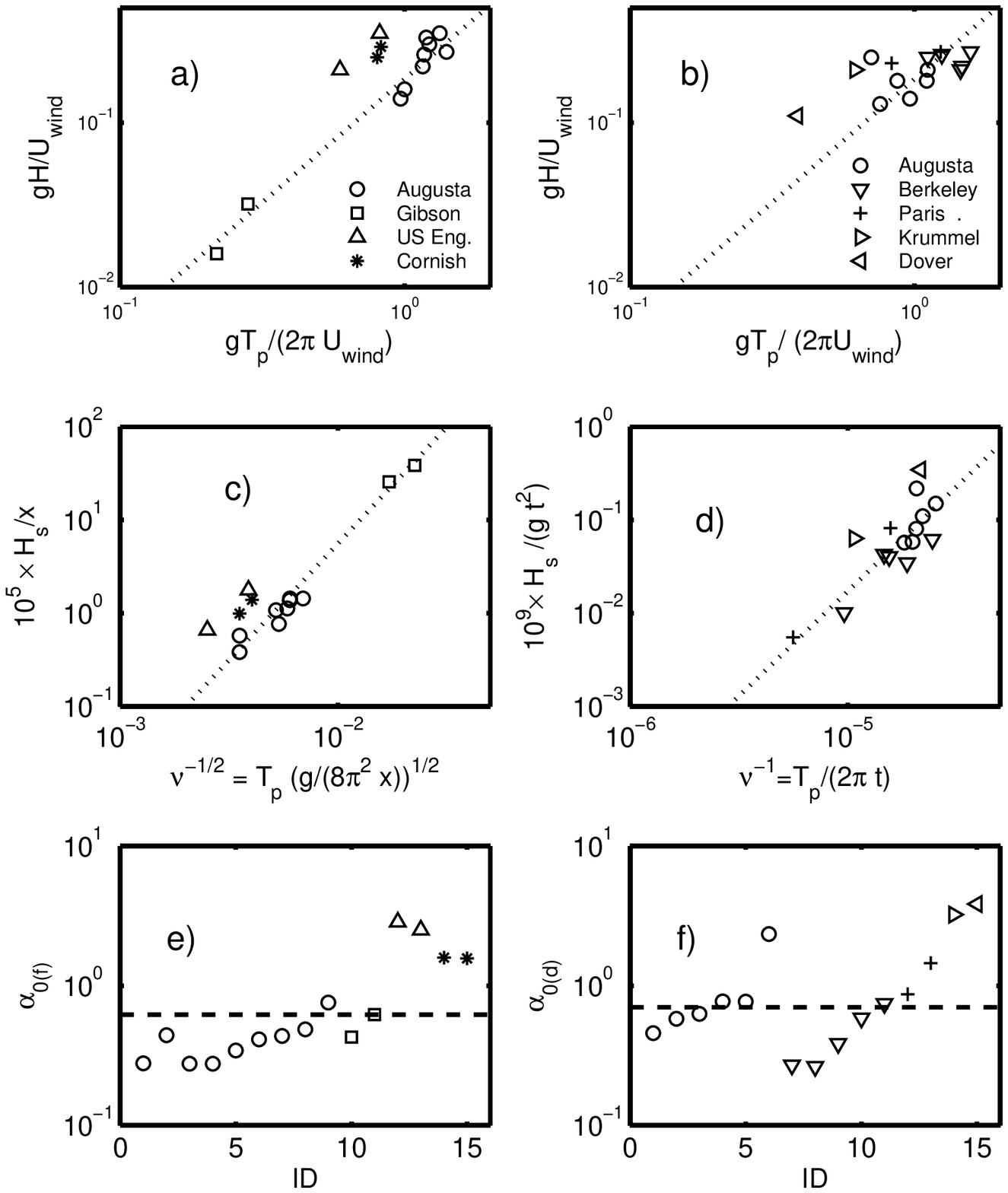}
  \caption{Data by \citet{SverdrupMunk1947}: upper row \emph{(a,b)} -- wave height-period dependencies within the conventional wind speed scaling, the dotted lines show the \citet{Toba1972} law; middle row  -- $\tilde H (\tilde T)$ scaled by fetch \emph{(c)} or time \emph{(d)}; lower row -- estimates of invariants $\alpha_{0(f)}$ and $\alpha_{0(d)}$. Left column -- data of Table II for fetch-limited cases, right -- Table III for duration-limited case from \citet{SverdrupMunk1947}. Observed ships' and authors' names are given in the legends of the upper panels.}\label{fig7}
\end{figure}
Our special attention to work by \citet{SverdrupMunk1947} is not limited by historical aspects of wind wave studies. The theoretical construction of this paper appears to be quite close to our physical model. Ten years before the formulation  of the spectral approach for wind waves \citet{SverdrupMunk1947} showed that \emph{`the concept of `significant waves' is essential for purpose of forecasing'}. They started their theory  from the equation of the integral balance of energy \citep[see eq.~47 in][]{SverdrupMunk1947}, i.e.  the counterpart of the second equation of our model (\ref{eq:Kin_split}). The effect of nonlinear interactions has been described by a semi-empirical relationship between two dimensionless parameters: wave steepness and wave age. The experimental data available at that time have been also used to specify parameters of the model.

Fig.~\ref{fig7} presents the data of Tables~II and III of Appendix II by \citet{SverdrupMunk1947}. Only measurements containing full triads of wave height, period and fetch (duration) were taken into account. The upper row in fig.~\ref{fig7}\emph{a} shows fetch- and duration-limited data in fig.~\ref{fig7}\emph{b}  made dimensionless by using the conventional wind speed scaling.  One can notice the rather high scatter of data points around  Toba's $3/2$ law (shown by dotted line). Strong deviations from Toba's law are associated with four particular observers: \emph{US Eng.} and \emph{Cornish }for fetch- and \emph{Kr\"{u}mmel}, \emph{Dover} for duration-limited observations.

The alternative wind-free scaling plots (\ref{def:HTndFetch},\ref{def:HTndDur}) in fig.~\ref{fig7}\emph{c,d} also show high discrepancies from the reference dependencies (\ref{eq:H2TDur},\ref{eq:H2TFetch}) but the ranges of the corresponding dimensionless wave heights and periods are wider, especially, for the duration-limited case (fig.~\ref{fig7}\emph{d}).  Again, we can notice a clear disparity of points associated with the source of data and the method of estimates of wave parameters. Data of \emph{USS Augusta} and \emph{Gibson } (fetch-limited) and of \emph{USS Augusta} and \emph{Berkeley} (duration-limited) show better correspondence to theoretical dependencies (dotted lines). Wave periods in these observations have been estimated directly while those of \emph{Kr\"{u}mmel} and \emph{Paris} have been computed from measurements of wavelength .

Data in fig.~\ref{fig7}\emph{c,d} cover the same range of dimensionless $\tilde T, \, \tilde H$ as simulations by \citet{ZRPB2012} in our fig.~\ref{fig4}. One point of duration-limited data (fig.~\ref{fig7}\emph{d}, \emph{Paris}) appears below the lower limit of $\tilde T$ in fig.~\ref{fig4}\emph{b}, i.e. in the range we treated as saturated wave field (see comments in sect.~\ref{sect:FetchSim}). As an opposite extreme, note the data of \emph{Gibson} that fit both  approaches (Toba's and the new one) fairly well (cf.~fig.~\ref{fig7}\emph{a,c}). Two points of \emph{Gibson} corresponds to very young waves in terms of wind speed scaling (wave age $C/U_{10}$ $0.22$ and $0.28$, see fig.~\ref{fig7}\emph{a}). Short fetches ($1.3$ and $3.5$ kilometers) put these point into the upper right of fig.~\ref{fig7}\emph{c}. Note, that in terms of `life distance' $x/\lambda$ these points correspond to approximately $152$ and $267$ wave periods.

The bottom raw, fig.~\ref{fig7}\emph{e,f} demonstrates surprising correspondence of experimental estimates of the invariant $\alpha_0$ to our theoretical values. Again, data of \emph{USS Augusta}, \emph{Gibson} and \emph{Berkeley} show the best fit to $\alpha_{0(d)}=0.7$ and $\alpha_{0(f)}=0.62$.

\subsection{Data of the Field Research Facility waveriders}
The data of waveriders in near-shore area look very attractive for illustrating the law (\ref{eq:key_result}) in terms of dependency of wave height on wave period (\ref{eq:H2TFetch}).
This simple $5/2$ power law dependency has been checked for data available  at the web-site of Field Research Facility of the US Army Corps of Engineers \verb"http://www.frf.usace.army.mil/". A summary  of the data is given in Table~\ref{table:2}. The waveriders collected data for many years, say, Buoy 630 -- since 1997 and Buoy 430 since 2008. Figures in Table~\ref{table:2} show the total number of measurements and the number of data fitting with a fetch-limited setup. We have selected records when offshore wind direction was $\pm 30^\circ $  from a direction perpendicular to the coast. For Buoy 200 there is a problem of the direction, it is sheltered by cape from the North, this is the closest coastline but there is almost no wind from this direction. Direction $280^\circ$ has been taken as an alternative one and the corresponding fetch has been set as $3$ times longer than one from the closest shoreline.
\begin{table}
\centering
\begin{tabular}{llccc}
  \hline
  Buoy ID & Fetch & Wave direction  & Total number &  Number of points\\
  & & (degrees) & of data points & of off-shore waves  \\
  \cline{1-5}\\
  190 & $6112 $ m & 320 &  23772 & 102 \\
  192 & $18500$ m & 260 &  20526 & 35 \\
  200 & $18500\,\, {\rm m}\times 3$ & 280 & 5773 & 24 \\
  430 & $18500$ m & 250 & 70995 & 606 \\
  630 & $3000$ m & 250 & 184195 & 171 \\
  \hline
\end{tabular}
\caption{Summary of waverider data of the Field Research Facility of the US Army Corps of Engineers. Wave direction is reported by the direction from which the waves originate, i.e. waves coming from the West have direction $270^\circ$.}
\label{table:2}
\end{table}

\begin{figure}
  \centering
 \includegraphics[scale=0.75]{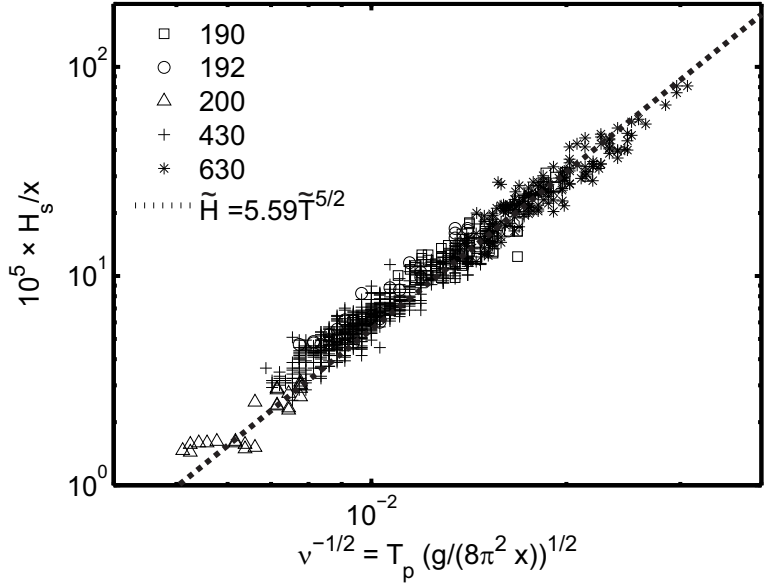}\\
  \caption{Data of FRF waveriders \emph{vs} theoretical dependence  (\ref{eq:H2TFetch}) shown as dotted line. Different buoy data are shown by different symbols (see legend and Table~\ref{table:2}).}\label{fig8}
\end{figure}

On-shore winds are dominating in this area and only $0.31$\% of total number of data (938 of 305261) have been qualified as consistent with a fetch-limited setup of wave growth. Being presented in non-dimensional variables (\ref{def:HTndFetch}) these data match the theoretical dependence $\tilde H(\tilde T) $ (\ref{eq:H2TFetch}) fairly well as seen in fig.~\ref{fig8}. The data of the buoys  cover a wide range of dimensionless periods (`life distances') similar to the figures above for simulations of wave growth and the historical data by \citet{SverdrupMunk1947} (cf.~figs.~\ref{fig4}--\ref{fig7}).
A slight overshoot relatively to the theoretical dependence can be explained by a systematic underestimating of wave periods. Underestimating of fetch can also contribute into this overshoot when the wave development is slowed down  by the coast sheltering effect. Anyway, fig.~\ref{fig8} is consistent with the new scaling to the classic problem of fetch-limited wind wave growth.

The scaling gives a good approximation for the problem of  wave growth at relatively small slanting fetches, in our present knowledge up to $40^\circ$ from the direction perpendicular to the coastline. At larger angles the wave growth off-shore is accompanied by  a complex system of propagating waves and along-shore modes as it was shown in recent simulations of the fetch-limited growth in which the  effect of nonlinear wave-wave interactions has been accounted for in  full \citep[e.g.][]{GagnaireThese,ZRPB2012}.

\section{Sea wave physics in wind wave tank experiments  }
According to common belief, dynamics of water waves observed in  wind-wave tanks differ dramatically from those of waves observed in open seas. All the experimental facilities are considered as too short for reaching full development of wave-wave interactions and for observing the related statistical properties of wind wave fields. At the same time, a number of laboratory results has been generalized successfully for  in-situ wave field conditions. The most famous example is the $3/2$ power law by \citet{Toba1972}  that links dimensionless wave height and wave period. The key physical scale of the law is friction velocity $u_*$  that quantifies the transport of momentum from the air flow to waves. For steady conditions, this parameter can be estimated from the vertical turbulent momentum flux in air using the relation: $u_*=\langle U^\prime W^\prime \rangle^{1/2}$.

\begin{figure}
  \centering
  \includegraphics[scale=0.6]{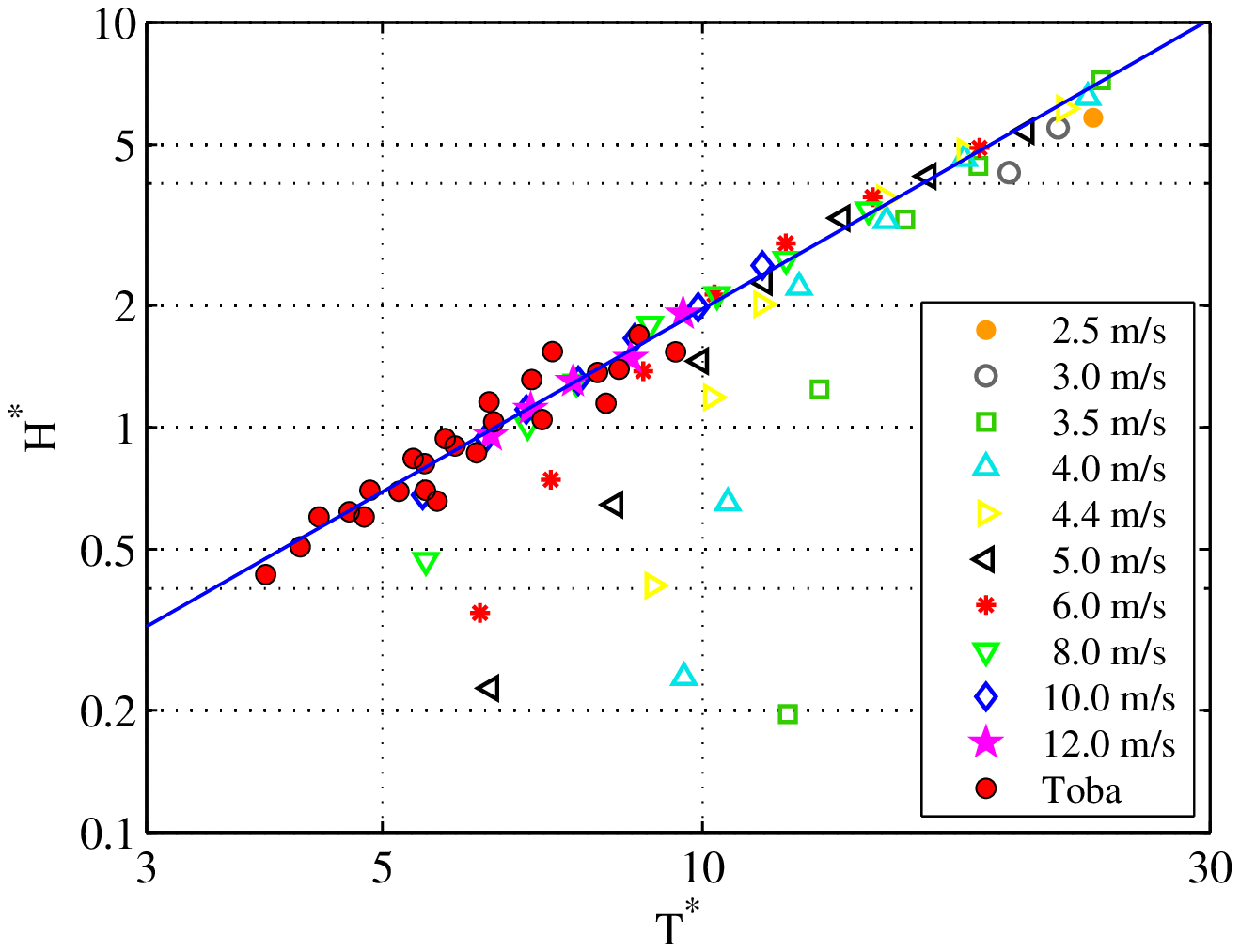}
    \includegraphics[scale=0.6]{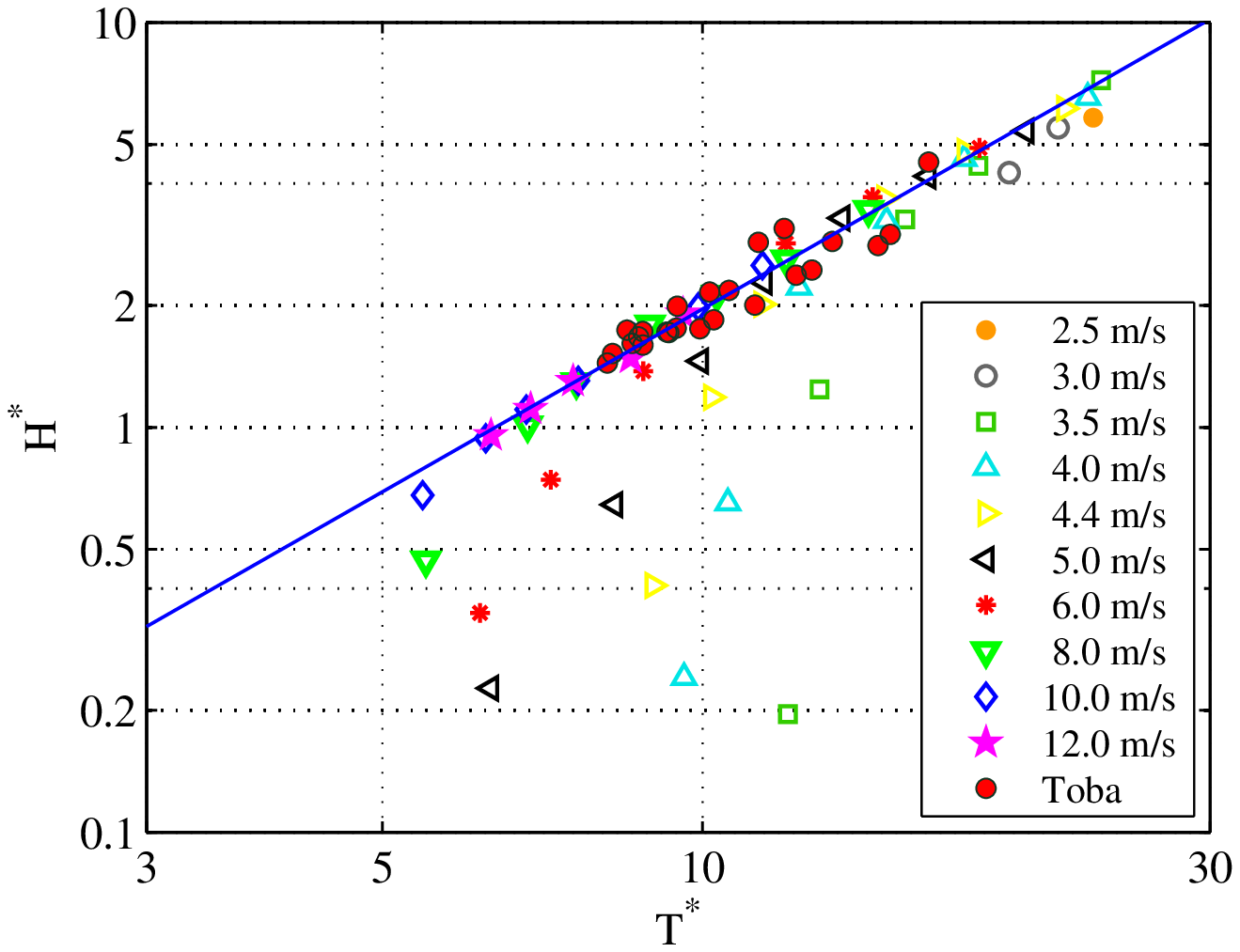}
    \caption{Data of wind-wave tank studies of wave growth by \citet{Toba1972} (filled symbols) and by \citet{Caulliez2014} (empty symbols) made dimensionless within conventional friction velocity $u_*=\langle U^\prime W^\prime \rangle^{1/2}$ scaling. {\it a)} -- Toba's data as given in Table~1 of \citet{Toba1972}; {\it b)} -- friction velocity $u_*$ in  \citet{Toba1972} data is replaced by estimates at the same reference wind speed and fetch made on the basis of Caulliez measurements in the large Marseille wind wave tank.}\label{fig9}
\end{figure}

Fig.~\ref{fig9}{\it a} shows the classical Toba's representation of wind wave growth in terms of the dimensionless wind-dependent wave parameters $H^*$ and  $ T^*$. Both wave data set observed in laboratory by Toba (filled symbols) and the more recent one obtained by  \citet{Caulliez2014} (open symbols) are plotted. These data have been collected for mostly similar reference wind speed conditions, between $5$ and $12$ m$\cdot$s$^{-1}$, but in quite different experimental facilities of respective length $21$ and $40$ m. In addition, the air tunnel above Toba's wave tank was just $50$ cm in height by $75$ cm in width and, evidently, the air flow might be  affected by walls. The friction velocity was determined from the mean velocity profiles measured at three fetches by small cup anemometers. When compared to Caulliez data, the original data reported by \citet{Toba1972} in Fig.~\ref{fig9}{\it a} show laboratory waves `unrealistically  young' exhibiting  $T^*$ abnormally small associated with inverse age in the range $20-35$, i.e. propagating $20-35$ times slower than the wind speed estimated at standard height $10$ meters (note that $T^*$ corresponds to wave age $C/u_*$ ). A thorough examination of the data set has shown such very small values of wave age are related to abnormally high values of friction velocity.

The observations made by Caulliez in the large wind-wave tank in Marseille concern a wide range of wind speeds and fetches, up to $26$ m, and are based on much more advanced technology. The air tunnel above the water surface was more than $1.5$ m in height and $3.2$ m in width. The friction velocity was determined from careful hot X-wire measurements of the vertical profiles of the turbulent momentum flux in air, a quantity found constant within the water surface boundary layer \citep{CaulliezEtAl2008}. In  this experiment, the inverse wave age varied between $4$ and $18$.

In Fig.~\ref{fig9}{\it a}  both data sets by Toba and Caulliez follow quite well the Toba  $3/2$ power law. However, the Toba data are strictly separated from the Caulliez ones. In addition, observations by Caulliez appear to be capable to catch the transitional stages of wind wave evolution at the entrance of the water tank while most of the values of the period $T^*$  obtained by Toba correspond to these transitional outliers of Caulliez data set.

Therefore, Toba's data have been re-analyzed by using more realistic values of wind parameters based on detailed measurements made in the large Marseille wind-wave tank. In brief, new friction velocity values have been derived from drag coefficient estimates obtained in Marseille at the same fetches and reference wind speeds as Toba's data.  Note, that these new values are in very good agrement with the $u_*$ values measured by \citet{Kawai1979} in a wind wave tank of comparable sizes as the tank used by \citet{Toba1961} in the experiments analysed here. In addition, significant wave height has been evaluated from mean wave height given by \citet{Toba1972} on the basis of the wave height probability distribution observed in Marseille and not from the Rayleigh distribution generally observed at sea. When used these new parameters for estimating $T^*$ and $H^*$, it is striking to see in fig.~\ref{fig9}{\it b} that both data sets then agree very well and fall into the same range of wave age. For well-developed wave fields at larger fetches, they also follow remarkably the $3/2$ power law.

\begin{figure}
  \centering
 \includegraphics[scale=0.6]{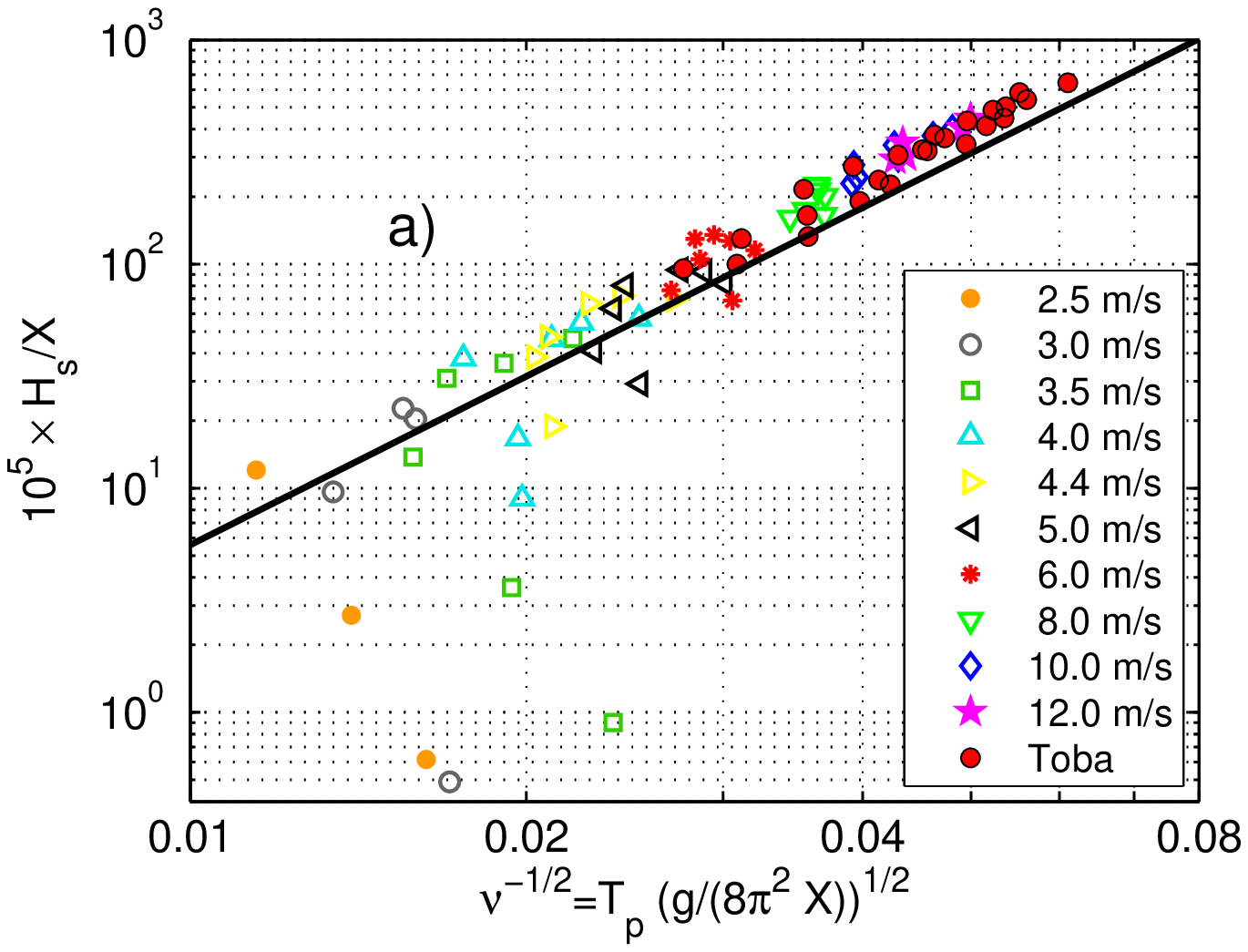}
  \includegraphics[scale=0.6]{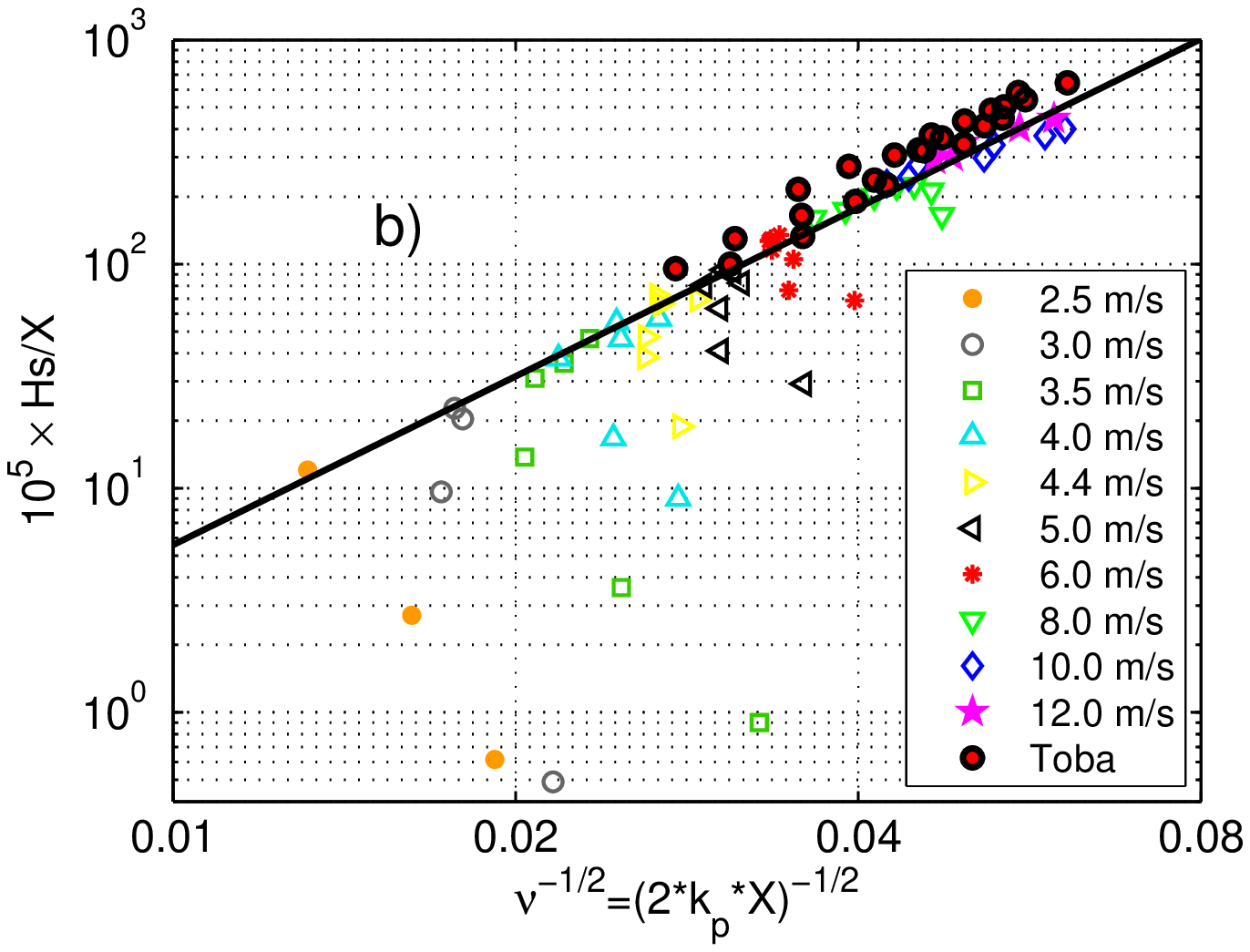}
    \caption{Data of wind-wave tank studies of wave growth by \citet{Toba1972} (filled symbols) and by  \citet{Caulliez2014} (empty symbols) in terms of  the new scaling (\ref{def:HTndFetch}). {\it a)} -- wave periods are estimated directly from time series of single-point wave height measurements; {\it b)} --  periods of wave development for  \citet{Caulliez2014} data are estimated  by taking into account not only gravity effects on wind wave dynamics but also capillarity and water surface drift current by considering a more realistic wave phase speed measured experimentally by using wave signals from a pair of gauges. These new estimates provide a better correspondence of the Caulliez data set to the new wave growth scaling derived theoretically.}\label{fig10}
\end{figure}

Figs.~\ref{fig10}  show the same data by G.~Caulliez and Y.~Toba plotted within the new scaling representation (\ref{def:HTndFetch}) based on the law of universality (\ref{eq:key_result}). In fig.~\ref{fig10}{\it a} significant wave height was estimated as four times the root-mean-square value of the water surface displacements for Marseille data and in the same way as in Fig.~\ref{fig9}{\it b} for Toba data while wave period is derived directly from spectral peak frequency.   The correspondence between both experimental  data sets and the theoretical power $5/2$ law dependence (\ref{eq:H2TFetch})  looks very good for well-developed wave fields observed at large fetches, but the  experimental points as a whole are located well above the theoretical curve given by (\ref{eq:H2TFetch}).

When noticed that the dimensionless wave period is a measure of the number of waves propagating over the fetch distance $x$, i.e. $\tilde T=(2|\kv_p| x)^{-1/2}$, and dynamics of wind wave fields observed in laboratory are governed not only by gravity but also by capillary forces and water surface flow drift, one can choose another scaling for evaluating wave period $\tilde T$. It consists of replacing the gravity $g$ by its counterpart in terms of phase velocity $C_p$ and wavenumber $|\kv|$ or wave period $T$, i.e. $g= C_p^2 |\kv_p|$ with $|\kv_p|=2 \pi/(C_p T)$, the phase velocity of dominant waves being estimated experimentally for Marseille data by using a cross-correlation method between two wave signals recorded by a pair of wave gauges \citep{Dudis1981}.

Fig.~\ref{fig10}{\it b} illustrates the advantage of this method for wind wave tank observations quite well: the agreement between the Caulliez data and the theoretical line improves strikingly and is undoubtedly much better than those of Toba data for which this correction could not be applied for lack of direct measurements of phase velocity. One can consider this fact as a good argument for supporting the appropriateness of the new scaling approach for describing wind wave fields observed in laboratory at large fetches (i.e. roughly above $6$ m fetch).

Finally, figs.~\ref{fig9},\ref{fig10} enable us to consider two issues thoroughly: the success of the new approach for wind wave tank observations and the related relevance of the conventional wind speed scaling within the Toba's law.  To support this viewpoint, one can also notice that in fig.~\ref{fig10}, the range of dimensionless periods $\tilde T$ is overlapping the data of Field Research Facility for buoys 190 and 630 located at the minimal distances from the coast, i.e. at 3 and 6 kilometers respectively (see figs.~\ref{fig8}). The equivalent lifetime in terms of wave periods for the wind wave tank data can be estimated at $50$ to $300$ periods, i.e. values quite close to  those obtained by \emph{Gibson} ($\tilde T$ is $152$ and $266$) mentioned in our analysis of \citet{SverdrupMunk1947} results. Thus, within the new approach, wind wave tanks of length above 10 m at least are large enough for modelling the nonlinear interactions of sea waves and statistical properties of the resulting wave field. Note, however, that from the conventional viewpoint of wind speed scaling the criterion of validity of wind wave tank observations for investigating wind-wave coupling is not so straightforward.

The data presented in this and the previous sections \citep[see figs.~\ref{fig7} for data by][]{SverdrupMunk1947} show quite good correspondence of the data both with our theoretical dependencies and Toba's law (especially, for wind wave tank data). It reflects, in a sense, the proximity of the basic physical assumptions of both the purely theoretical model of this paper and the theoretico-empirical model by \citet{Toba1972}. Both models consider physical mechanisms providing a local balance \citep[as emphasized in the title of the paper by][]{Toba1972}. Within the Toba model the local balance  of wind stress and wave momentum implies  a saturation of wave state, i.e. an immediate relaxation to the balance. Such balance gives the famous  exponent $3/2$ of height-to-period dependence.

Within our model the strong nonlinear relaxation provides a universal shaping of wave spectra and, then, the universality features of  the balance of integral quantities (total net input and growth rate). This universality is expressed, first, by a continuum of exponents of wind wave growth, i.e. by a continuum of wind-wave coupling scenarios in contrast to the fixed $3/2$ by Toba. Secondly, it gives  the universal wind-free invariant (\ref{eq:key_result}) that reflects additional links of pre-exponents of self-similar solutions, again, in contrast to the \citet{Toba1972} constant $B$ of which the universality is quite questionable. The absence of universality of `wind-based' models of wave growth is explained naturally by complexity of wind wave coupling when air turbulence, stratification etc. break the universality. The proximity of the `wind-based' model by Toba and our `wind-free' model, especially in the wind wave tank experiments (cf.~figs.~\ref{fig9} and \ref{fig10}), is likely dealing with a sort of standard conditions of wind-wave coupling in the tanks.

%
\section{Discussion and conclusions}
Presenting a new theoretical concept of wind wave growth, this paper revisits a variety of previous theoretical, numerical and experimental results. This final section aims at showing consistency and logics of this reconsideration.

\subsection{Theory}
The key theoretical result of this paper is  the invariant (\ref{eq:key_result}) of wind wave growth that looks paradoxical within the conventional physics of wind waves. The basic assumption made on the  dominating role of nonlinear wave transfer  leads to the property of self-similarity of wave spectra for the reference cases of duration- and fetch-limited wave growth. The parameters of these self-similar solutions are rigidly linked: the exponents $p,\,q$ obey the linear links (\ref{eq:p2qdur}, \ref{eq:p2qfetch}) while pre-exponents in (\ref{eq:exp_law}\textit{a,b}) satisfy a sophisticated nonlinear relationship between wave energy and flux of this energy (total wave input, e.g. eq.~\ref{eq:WTGR}).

The self-similar solutions can be written in the form of an invariant. This invariant looks paradoxical, first of all, because it is not determined by the initial conditions (pre-exponent of the power-law solution). The system `forgets' its initial state completely but still `remembers' the rate of wave growth (exponents $p_\tau\,(p_\chi)$). This remarkable result is mathematically strict for the families of self-similar solutions.

The next step consists of a physical assumption of the quasi-universality of the wave spectrum shapes \citep[invariance of spectral shaping in the words of][]{Hass_ross_muller_sell76}. It makes the invariant (\ref{eq:key_result}) to be independent on wave growth rate $p_\tau\,(p_\chi)$ -- the parameters of families of self-similar solutions. One can treat this result as an adiabatic relationship that describes the whole family of solutions. Again, the result is not trivial: the adiabatic invariant does not refer to a parameter of adiabaticity.

In fact, the most striking paradox of our study is the invariant itself of \emph{wind wave growth} that \emph{does not depend explicitly on wind parameters} (say, conventional wind speed $U_{10}$ or $u_*$). The fact of the absence of wind speed in our analysis is really surprising. It means that the effect of wind forcing is recorded by inherent parameters of the wave field, in particular, the wave steepness $\mu$ and instant number of waves $\nu$. \\

\begin{center}
\emph{Waves  chronicle wind} rather than \emph{wind rules waves}.
\end{center}

\vskip 7pt
\noindent The absence of  wind speed dependency leads to a new scaling of dimensionless wave height and period (\ref{eq:H2TDur}, \ref{eq:H2TFetch}): simple time or fetch become key physical scales irrespectively to conditions of wind forcing. We show that the new theoretical dependencies consist with conventional theoretico-empirical ones fairly well \citep[e.g.][]{Hass_ross_muller_sell76,Carter1982}. Exclusion of wind speed from these dependencies gives discrete exponents $9/4,\,5/2$ and, more surprisingly obtains consistent estimates of basic constants $\alpha_{0(d)},\,\alpha_{0(f)}$.

Our reference to the theoretico-empirical model by \citet{Hass_ross_muller_sell76} is quite representative. Our purely theoretical approach shows that the empirical estimations of the parameters in the JONSWAP spectrum and the particular shape of this spectrum are exhaustive: self-similar solutions and their adiabatic extension can be substantiated  by much more general physical principles.

\subsection{Simulations of wind wave growth}
 We do not present our numerical results as a proof of the theory. Moreover, we stress, that all the experiments have been accomplished almost 10 years ago with no reference to our findings in this paper. Few runs for duration-limited setup from \citet{BPRZ2005,BBRZ2007} have been taken arbitrary in order to show very good agreement  with predictions of the new  theory.

Recent numerical study of fetch-limited growth by \citet{ZRPB2012} gave us a broader view on the classic problem of wave growth. We found that the fetch-limited regime is just a stage  in these simulations where wave growth is limited in space.  When the waves from the coast arrive at an opposite side of the simulation domain the waves continue to grow in a duration-limited scenario.

\subsection{Field experiments on wind wave growth}
Field studies of wave growth give us a major part of support for the new theory. First, we show that experimental approximations imply quite often links between parameters of power-law fits (\ref{eq:exp_law}). Parameterizations by \citet{HwangWang2004} (\ref{eq:p2qPaulDur}, \ref{eq:p2qPaulFetch}) reproduce the theoretical links (\ref{eq:p2qdur}, \ref{eq:p2qfetch}) fairly well: minor quantitative difference makes the invariant (\ref{eq:key_result}) to be weakly dependent on dimensionless fetch (or wave age).

An outcome of our historical review of the brilliant paper by \citet{SverdrupMunk1947} is two-fold. First, we show that wave observations made during   World War II are consistent with our theory. Secondly, we emphasize the parallel  of our self-similarity approach and the concept of significant wave height. Both approaches  represent the wave field as one described by a minimal number of parameters: significant wave height  and peak period in contrast to more detailed but much more expensive description of wave field as an ensemble of a great number of spectral components.

Data of wave riders of the East Coast of the US are presented as an impressive justification of the new theoretical outcome. Wave heights and periods scaled by the new wind speed free way show remarkable closeness to the law of $5/2$ for the fetch-limited case.  One has no doubt that more experimental proofs of the validity of the approach can be found in data of wave riders all over the World Ocean.

\subsection{Wave tank experiments: beyond the formal validity of the statistical description?}
Results of wave tank experiments are very representative. First, the wave tank experiments are generally considered as `too far' from wind sea reality. We show that it is likely not the case. In terms of the new wind free scaling the results of experiments in the Wind Research Facility in Marseille fit theoretical dependencies remarkably well. They correspond to high values of dimensionless $\tilde H, \tilde T$, i.e. to relatively short fetches of just a few dozens of instant wavelengths. Expectedly, the statistical description used in this work may not apply  in this case. At the same time, the range of the wave tank experiments overlaps the one of the wave riders presented in sect.~4 in terms of the number of waves within the fetch. It offers good perspectives for modelling sea waves in  wind-wave facilities \citep{ZavadskyShemer2013}.

\begin{acknowledgments}
This work is sponsored by the Office of Naval Research (Naval Research Laboratory PE 61153N),   ONR grant N000141010991 and Centre National d'Etudes Spatiales (TOSCA/SMOS-Ocean project). V.E. Zakharov gratefully acknowledges support of Russian Science Foundation N14-22-00174.  S.I. Badulin has been supported by Russian Foundation for Basic Research  14-05-00479-a and  grant of the Government of the Russian Federation 11.G34.31.0035. Authors are thankful for data provided by the Field Research Facility, Field Data Collections and Analysis Branch,  US Army Corps of Engineers, Duck, North Carolina. The NRL publication number is NRL/JA/7260—14-0232.
\end{acknowledgments}

\appendix
\section{Self-similar solutions for growing wind seas}

\subsection{Duration-limited case}
Let us consider the duration-limited case, i.e. a spatially homogeneous case when $\partial E/\partial x\equiv \partial E/\partial y\equiv 0$. Solutions for the conservative kinetic equation (\ref{eq:Kin_split}{\it a}) with homogeneity condition (\ref{eq:Homo}) can be found in the form of incomplete  self-similarity  (\ref{eq:SSadimtext}).

After substituting (\ref{eq:SSadimtext}) into (\ref{eq:Kin_split}{\it a}) one has
\begin{equation}\label{eq:ss_detail}
  (p_\tau+4q_\tau)\Phi_{p_\tau}(\xiv)+2q_\tau \xiv \nabla_\xi \Phi_{p_\tau}=a_\tau^2 b_\tau^{-17/2} \tau^{R} S_{nl}[\Phi_{p_\tau}(\xi)]
\end{equation}
where exponent $R=2p_\tau-9 q_\tau+1$ should be zero to cancel the explicit dependence on time $\tau$ and to leave a dependence on self-similar argument $\xiv$ only. It gives the linear link (\ref{eq:p2qdur}) between exponents $p_\tau$ and $q_\tau$.
An additional link of coefficients $a_\tau$ and $b_\tau$ (\ref{eq:a2bdur}) can be introduced by simple re-scaling of dimensionless variables.
The total dimensionless energy for solutions (\ref{eq:SSadimtext}) with (\ref{eq:a2bdur}) becomes
\begin{equation}\label{eq:totEn}
  \tilde E_{tot}=\int\int_{-\infty}^{+\infty} a_\tau \tau^{p_\tau+4q_\tau} \Phi_{p_\tau}(\xiv) d\kv=a_{\tau}^{9/17} \tau^{p_\tau} I_{\tau}
\end{equation}
where
\begin{equation}
\label{def:Id}
I_\tau=\int \int_{-\infty}^{+\infty}\Phi_{p_\tau}(\xiv) d\xiv
\end{equation}
Links (\ref{eq:p2qdur}, \ref{eq:a2bdur})  are of key importance for further consideration. First, equation (\ref{eq:ss_detail}) for shape function $\Phi_{p_\tau}(\xiv)$ with (\ref{eq:a2bdur}) depends on exponents $p_\tau,\,q_\tau$ but  appears to be independent on coefficients $a_\tau,\,b_\tau$. Secondly, the self-similar solutions (\ref{eq:SSadimtext}) with the integral (\ref{eq:totEn}) depending on two parameters only (say, $a_\tau$ and $p_\tau$) are consistent  with a power-law dependence of net wave input on time in (\ref{eq:Kin_split}{\it b}).

A characteristic frequency $\omega_*$ can be introduced in different ways for a given spectral shape function $\Phi_{p_\tau}(\xiv)$. Mean-over-spectrum frequency is written as follows
\begin{equation}
\label{def:omegamean}
\tilde \omega_m=\frac{\int \int_{-\infty}^{+\infty}\tilde \omega \Phi_{p_\tau}(\xiv) d \xiv}{\int \int_{-\infty}^{+\infty} \Phi_{p_\tau}(\xiv) d \xiv}=a_\tau^{-2/17}\tau^{-q_\tau} {J_\tau}\cdot{I_\tau^{-1}}
\end{equation}
where
\begin{equation}\label{def:Jd}
  J_\tau=\int \int_{-\infty}^{+\infty}|{ \xiv} | \Phi_{p_\tau}(\xiv) d \xiv.
\end{equation}
Peak frequency $\tilde \omega_p$ that corresponds to a maximum of the shape function $\Phi_{p_\tau}(\xiv)$, evidently, have similar dependencies on time and parameter $a_\tau$
\begin{equation}
\label{def:omegap}
\tilde \omega_p = h_{p_\tau} \tilde \omega_m = h_{p_\tau} a_\tau^{-2/17}\tau^{-q_\tau} J_\tau\cdot I_\tau^{-1}
\end{equation}
where the coefficient $h_{p_\tau} < 1$ for wind-wave spectra -- mean frequency is generally higher than the peak one. Note, that this coefficient depends on exponent $p_\tau$. Below we use the spectral peak frequency $\omega_p$ unless otherwise stated.

While total wave energy (\ref{eq:totEn}) and  characteristic frequency (\ref{def:omegamean}, \ref{def:omegap}) are power law functions which exponents are linked by linear relationship (\ref{eq:p2qdur}) one can construct easily a time-independent invariant in the form
\begin{equation}\label{eq:key_alt}
  \tilde E_{tot}^s \tilde \omega_p^y \tau = a_{\tau}^{9s/17-2y/17}\tau^{sp_\tau-yq_\tau+1} I_\tau^{s-y} J_\tau^{y} h_\tau^y
\end{equation}
by choosing proper exponents $s$ and $y$. Exponents $s=2$ and $y=9$  cancel dependence on time $\tau$ in (\ref{eq:key_alt}). The remarkable fact is that this choice cancels dependence on parameter $a_\tau$ as well. The `magic link' (\ref{eq:p2qdur}) between exponents $p_\tau$ and $q_\tau$ gives a time-independent invariant that depends on one parameter only of the family of self-similar solutions (\ref{eq:SSadimtext}). The invariant (\ref{eq:key_alt}) can be associated with the weakly turbulent law of wind-wave growth by \citet{BBRZ2007} in the form of the Kolmogorov relationship between energy and energy flux (total net input) (\ref{eq:WTGR}).
Finally, one has the invariant in a remarkably concise and physically transparent form in terms of wave steepness $\mu$ (\ref{def:steepness}) and wave lifetime $\nu$ as defined by (\ref{eq:nu})
\begin{equation}\label{eq:invNew}
  \mu^4 \nu = I_\tau^{-7} J_\tau^{9} h_\tau^9 = \alpha_{ss(d)}^3 p_\tau =\alpha_{0(d)}
\end{equation}
Here we use subscript $(d)$ for  duration-limited case.

\subsection{Fetch-limited case}
Self-similar solutions for fetch-limited setup can be considered quite similarly to the duration-limited case. Assuming wave field to be stationary ($\partial E/\partial t\equiv 0$) and growing in coordinate $x$ one has self-similar solutions  in the form (\ref{eq:SSadimfetch}) and `magic links' (\ref{a2bfetch}) quite similarly to the duration-limited case.

For dimensionless energy (cf.~\ref{eq:totEn}) with (\ref{a2bfetch}) one has
\begin{equation}\label{eq:totEnfetch}
  \tilde E_{tot}=\int\int_{-\infty}^{+\infty} a_\chi \tau^{p_\chi+4q_\chi} \Phi_{p_\chi}(\zetav) d\kv=a_{\chi}^{5/9} \chi^{p_\chi} I_\chi
\end{equation}
where
\begin{equation}
\label{def:If}
I_{\chi}=\int \int_{-\infty}^{+\infty}\Phi_{p_\chi}(\zetav) d\zetav
\end{equation}
Mean-over-spectrum frequency is written as follows
\begin{equation}
\label{def:omegameanfetch}
\tilde \omega_m=\frac{\int \int_{-\infty}^{+\infty}\tilde \omega \Phi_{p_\chi}(\zetav) d \zetav}{\int \int_{-\infty}^{+\infty} \Phi_{p_\chi}(\zetav) d \zetav}=a_\chi^{-1/9}\chi^{-q_\chi} {J_\chi}\cdot{I_\chi^{-1}}
\end{equation}
and the peak one
\begin{equation}
\label{def:omegapfetch}
\tilde \omega_p = h_{p_\chi} \tilde \omega_m = h_{p_\chi} a_\chi^{-1/9}\chi^{-q_\chi}J_\chi\cdot I_\chi^{-1}
\end{equation}
where
\begin{equation}\label{def:Jf}
  J_\chi=\int \int_{-\infty}^{+\infty}|{ \zetav} | \Phi_{p_\chi}(\zetav) d \zetav.
\end{equation}
The fetch-independent invariant can be derived in the same way as the one for duration-limited setup (\ref{eq:key_alt}) and related with formulations by \citet{BBRZ2007,GBB2011}
\begin{equation}\label{eq:GrowthLawfetch}
  \tilde E_{tot} \tilde \omega_p^4=\alpha_{ss(f)}\left(\tilde \frac{\omega_p^2}{2} \frac{\partial \tilde E_{tot}}{\partial  \chi}\right)^{1/3}
\end{equation}
Finally, one has the invariant in terms of wave steepness $\mu$ (\ref{def:steepness}) and wave lifetime -- life distance
\[
\nu=2\kv_p \xv
\]
which is identical to the duration-limited case
\begin{equation}\label{eq:invNewfetch}
  \mu^4 \nu = I_\chi^{-8} J_\chi^{10} h_\chi^{10} = \alpha_{ss(f)}^3 p_\chi = \alpha_{0(f)}.
\end{equation}
The right-hand sides of (\ref{eq:invNew},\ref{eq:invNewfetch}) are formally different and are determined by self-similar functions $\Phi_\tau(\xiv),\, \Phi_\chi(\zetav)$.



\end{document}